\documentclass[useAMS,usenatbib]{mn2e}
\usepackage{graphicx,amssymb, txfonts,pstricks}

\def\pb{Pa$\beta$}
\def\br{Br$\gamma$}
\def\ha{H$\alpha$}
\def\hb{H$\beta$}
\def\feii{[Fe\,{\sc ii}]}
\def\pii{[P\,{\sc ii}]}
\def\oiii{[O\,{\sc iii}]}

\def\nii{[N\,{\sc ii}]}

\def\h2{H$_2$}
\def\p1{Paper~I}

\title[Near-IR dust and line emission from Mrk\,1066]{Near-IR dust and line emission from the central region of Mrk\,1066: 
Constraints from Gemini NIFS}
\author[Riffel, Storchi-Bergmann \& Nagar]{Rogemar. A. Riffel$^{1,2}$\thanks{E-mail:
rogemar@smail.ufsm.br}, Thaisa Storchi-Bergmann$^{2}$ and Neil M. Nagar$^{3}$
\\%\footnotemark[1]\thanks{This file has been amended to
 $^{1}$Universidade Federal de Santa Maria, Departamento de F\'\i sica, Centro de Ci\^encias Naturais e Exatas, 97105-900, Santa Maria, RS, Brazil \\ 
$^{2}$Universidade Federal do Rio Grande do Sul, Departamento de Astronomia, Instituto de F\'\i sica, CP 15051, 91501-970, Porto Alegre, RS, Brazil\\
$^{3}$Astronomy Group, Departamento de F\'\i sica, Universidad de Concepci\'on, Casilla 160-C, Concepci\'on, Chile}

\begin{document}

\date{Accepted 1988 December 15. Received 1988 December 14; in original form 1988 October 11}

\pagerange{\pageref{firstpage}--\pageref{lastpage}} \pubyear{2002}

\maketitle

\label{firstpage}

\begin{abstract}

We present integral field spectroscopy of the inner $700\times700$\,pc$^2$ of the Seyfert galaxy Mrk\,1066
obtained with Gemini's Near-Infrared Integral Field Spectrograph (NIFS)
at a spatial resolution of  $\approx$35\,pc. This high spatial resolution allowed us to observe, for the first time
in this galaxy, an unresolved dust concentration with mass $\sim1.4\times10^{-2}\,{\rm M_\odot}$. This unresolved
concentration, with emission well reproduced by a blackbody with temperature $\sim$830\,K, is 
possibly part of the nuclear dusty torus. 
We compared maps of emission-line flux distributions and ratios with
a 3.6\,cm radio-continuum image and \oiii\ image in order to investigate the origin of the near-infrared emission.
The emission-line fluxes are elongated in PA$=135/315^\circ$ in agreement with the \oiii\ and radio images and, except for the H lines, are brighter
to the north-west than to the south-east. This close association with the radio hot spot implies that at
least part of the emitting gas is co-spatial with the radio outflow. The H emission is stronger to the south-east, 
where we find a large region of star-formation. The strong correlation between the radio emission and the highest 
emission-line fluxes indicates that the radio jet plays a fundamental role at these intensity levels. At lower emission-line fluxes this correlation disappears suggesting a
contribution from the plane of the galaxy to the observed emission.
% The \feii~and \pii\ flux distributions are elongated following the radio and \oiii\ flux distributions peaking at the same location at
% 1$^{\prime\prime}$ north-west of the nucleus, while the H recombination lines and H$_2$ lines peak at the nucleus.
The \h2\ flux is more uniformly distributed and has an excitation temperature of
$\approx2100$\,K. Its origin appears to be circum-nuclear gas heated by X-rays from the central active nucleus. The \feii\
emission also is consistent with X-ray heating, but its spatial correlation with the radio
jet and \oiii\ emission indicates additional emission due to excitation and/or abundance changes caused by shocks in the radio jet.
The coronal-line emission of [Ca\,{\sc viii}] and [S\,{\sc ix}] are unresolved by our observations indicating a distribution within 18\,pc from the nucleus.
 The reddening map obtained via the \pb/\br~line ratio ranges from $E(B-V)\approx0$ to $E(B-V)\approx1.7$ with the highest values defining a
 S-shaped structure along  PA$\approx$135/315$^\circ$. The emission-line ratios are Seyfert-like within the ionization cone indicating that
 the line emission is powered by the central active nucleus in these locations. Low ionization regions are observed away from the ionization cone, and may be powered 
by the diffuse radiation field which filters through the ionization cone walls. Two regions at  0\farcs5 south-east
 and at 1$^{\prime\prime}$ north-west of the nucleus show starburst-like line ratios, co-spatial with an enhancement in the emission of the
H lines. We attribute this change to additional emission from star forming regions.
  The mass of ionized gas  is $M_{HII}\approx1.7\times10^7\,{\rm M_\odot}$ and that of hot molecular
gas is $M_{H_2}\approx3.3\times10^3\,{\rm M_\odot}$.

\end{abstract}

%________________________________________________________________

\section{Introduction} \label{intro}

\begin{figure*}
\centering
\includegraphics[scale=0.85]{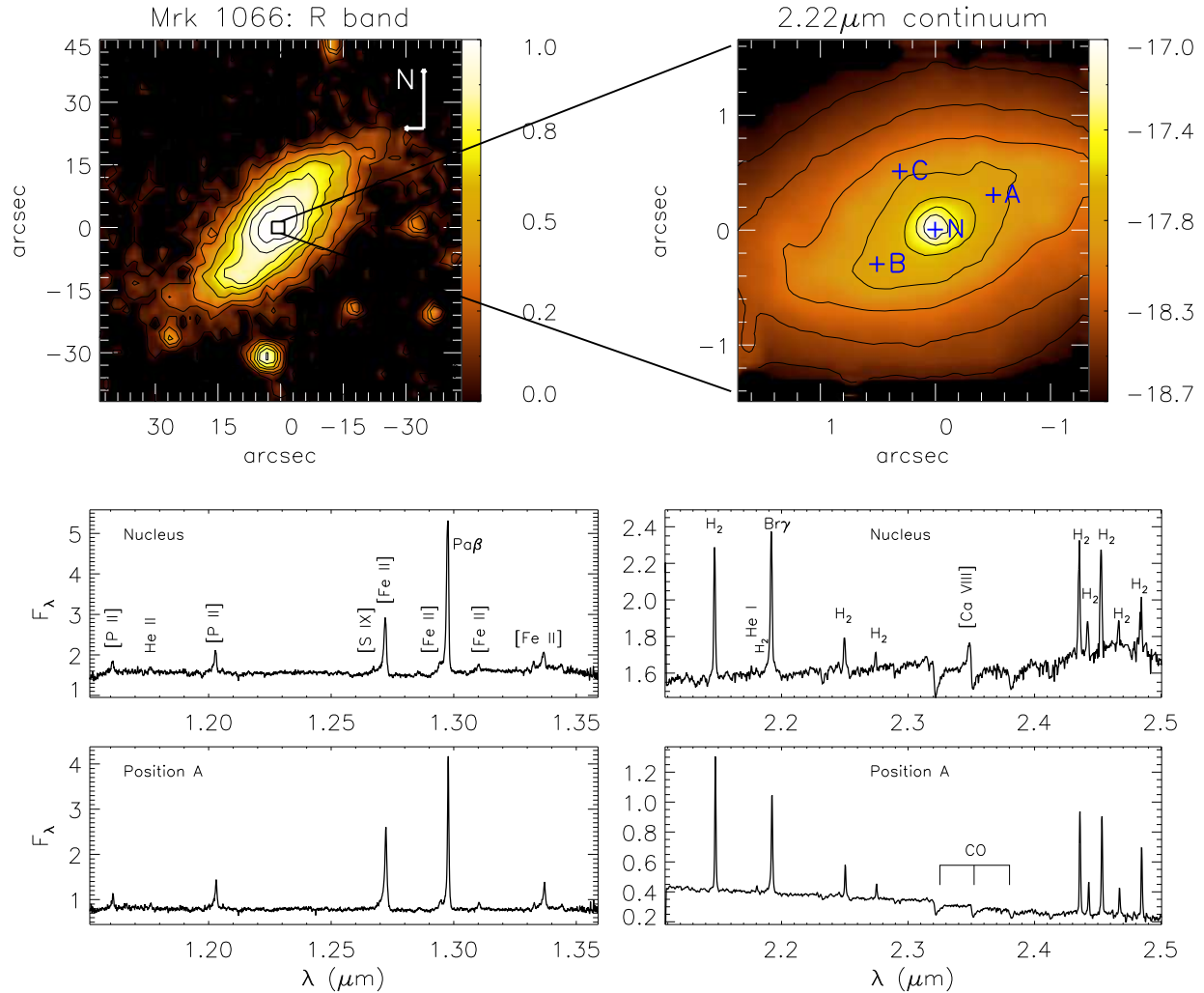}
\caption{Top left: Large scale image of Mrk\,1066 obtained at the R-band with the 1.22\,m Oschin Telescope on Palomar Mountain \citep{lasker96}.
 The central box shows NIFS field of view. Top right: $2.22\,\mu$m continuum map from NIFS observations. Bottom: Spectra for the nuclear position (N at top-right panel) 
and for the position A (0\farcs5\,north-west from the nucleus) for the J and K$_{\rm l}$-bands for an aperture of 0\farcs25$\times$\farcs25.
 The emission lines are identified at the nuclear spectrum and in the extra-nuclear spectrum we mark the K-band CO absorption band heads.}
\label{espectro}
\end{figure*}

The study of the gas distribution and excitation in the Narrow Line Region (NLR) of active galaxies has
 a fundamental importance in the understanding of the physical phenomena in the vicinity of the super-massive black holes (SMBH) in their 
centers. The excitation of the inner NLR can reveal how the radiation and mass outflows from the active galactic nucleus (AGN)
 interact with the surrounding inter-stellar 
medium (ISM). Several studies have been aimed to investigate the line emission from the NLR gas
 \citep[e.g.][]{rodriguez-ardila04,rodriguez-ardila05a,riffel06,riffel08,riffel09a,stoklasova09,storchi-bergmann09,barbosa09}. 
Nevertheless, its physics is still not completely understood.  

Besides the line emission, the nature of the nuclear continuum emission is an important issue in the study of
 AGNs. In particular, the study  of near-infrared (hereafter near-IR)  continuum  
can be used to probe the emission from hot dust located in the putative torus postulated by the Unified Model \citep{antonucci93}. 
Several studies revealed the presence of hot dust with temperatures of $T\sim900-1500\,$K and masses of
 $M_{\rm HD}\sim10^{-6}-10^{-2}\,{\rm M_\odot}$ \citep[e.g.][]{rieke81,rodriguez-ardila05b,rodriguez-ardila06,riffel09a,riffel09b,rogerio09}.  The 
high spatial resolution provided by adaptive optics spectroscopy in large telescopes can be used to constraint
 the location of the near-IR emitting dust \citep[e.g.][]{riffel09b}.

This work is aimed to investigate the near-IR emission-line flux distributions, gas excitation and continuum emission within the inner kiloparsec of 
Mrk\,1066 from spectroscopy obtained with the Gemini's Near-Infrared Integral Field Spectrograph \citep[NIFS,][]{mcgregor03}. 
In a following paper we discuss the kinematics of both the gas and stars. 
Mrk\,1066 was selected for this study because it presents extended radio \citep[e.g][]{ulvestad89,nagar99} and  near-IR line emission 
 \citep[e.g.][]{rodriguez-ardila05a,knop01} allowing us to investigate the role of the radio jet in the emission of the NLR gas. 
In addition, its central kpc has a high extinction in the optical \citep[e.g.][]{stoklasova09} and thus a near-IR study is better suited for the 
investigation of the gas excitation and kinematics.

\begin{figure*} % Second figure of results
\centering
\includegraphics[scale=0.85]{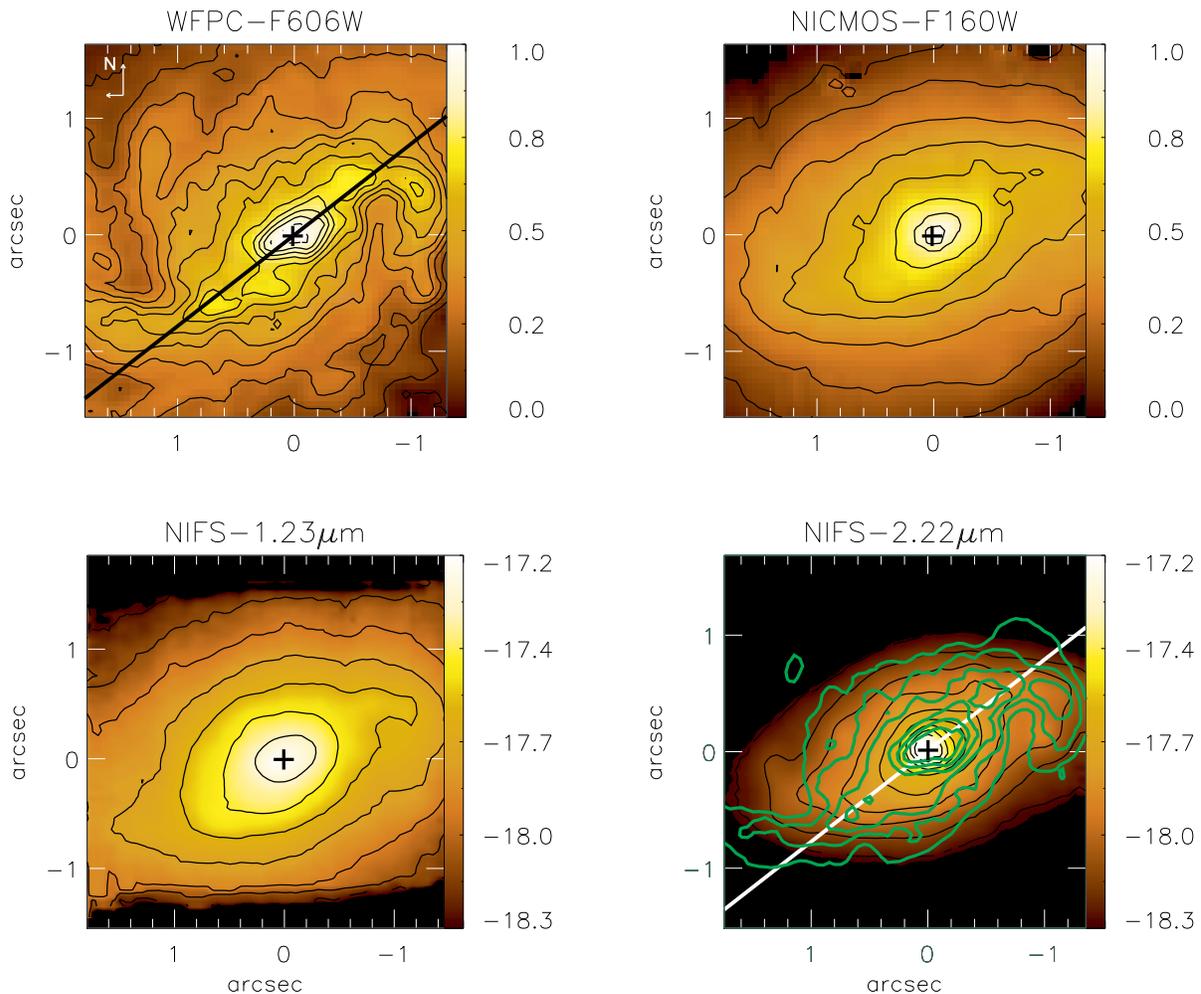}
\caption{Continuum images of the central 3$^{\prime\prime}\times$3$^{\prime\prime}$ of Mrk\,1066. Top left: Optical continuum image 
obtained with the HST-WFPC2 using the filter F606W  \citep{malkan98}. Top right: H-band continuum image obtained with 
HST-NICMOS with filter F160W. Bottom: Continuum images centered at 1.23$\,\mu$m (left) and  2.22$\,\mu$m (right) 
from NIFS observations. The HST images are in arbitrary units and NIFS images are in logarithmic flux units (erg\,s$^{-1}$\,cm$^{-2}$ \AA$^{-1}$). The straight
line shows the orientation of the major axis of the galaxy and the thick contours overlaid to the K-band map are from the optical continuum image.}
\label{cnt}
\end{figure*}

Mrk\,1066 is an SBO+ galaxy harboring a Seyfert 2 nucleus; it is  located at a distance of 48.6\,Mpc\footnote{Distance quoted in NASA/IPAC Extragalactic Database (NED -- 
http://nedwww.ipac.caltech.edu)}, for which 1$^{\prime\prime}$ 
corresponds to 235\,pc at the galaxy. {\it Hubble Space Telescope} (HST) narrow-band images show a ``jetlike'' feature in the \oiii$+$\hb~emission 
extending up to 1\farcs4 north-west from the nucleus  along the position angle PA=315$^\circ$, while the \ha+\nii~image is 
extended to both sides of the nucleus \citep{bower95}. Radio continuum images of Mrk\,1066, at 3.6, 6 and 20\,cm, show extended 
emission up to 1\farcs5 to both sides of the nucleus oriented approximately along the same PA of the optical line emission 
\citep{ulvestad89,nagar99}. Long slit spectroscopy shows that the near-IR emission lines are extended up to  5$^{\prime\prime}$  from the nucleus 
along the PA=135/315$^\circ$ with  distinct flux distributions for \feii, \h2~and H recombination lines suggesting different 
emission processes \citep{knop01}.

 This paper is organized as follows: In Section~2 we present the description of the observations and data reduction. In Sec.~3 we present the results 
for the continuum and line emission, as well as line-ratio maps. The discussion of the results  
is presented in Sec.~4 and the conclusions are presented in Sec.~5.

\section{Observations and Data Reduction}

\subsection{The Gemini NIFS data}
The Integral Field Unit (IFU) spectroscopic data of Mrk\,1066 were obtained with Gemini NIFS \citep{mcgregor03} 
operating with the Gemini North Adaptive Optics system ALTAIR in September 2008 under the programme GN-2008B-Q-30. The NIFS has a 
square field of view of $\approx3\farcs0\times3\farcs0$, divided into 29 slices with an 
angular sampling of 0$\farcs$103$\times$0$\farcs$042, optimized for use with ALTAIR. 

The observing procedure followed the standard Object-Sky-Sky-Object dither sequence, with off-source
sky positions since the target is extended, and individual exposure times of
600\,s. Two set of observations were obtained at different spectral ranges: the first at the J-band, centered at 
1.25\,$\mu$m and covering the spectral region from 1.15\,$\mu$m to 1.36\,$\mu$m, and the second at the K$_{\rm l}$-band, centered at 2.3\,$\mu$m 
covering the spectral range from  2.10$\,\mu$m to 2.26$\,\mu$m. At the J-band, the selected instrument configuration was
 the J\_G5603 grating and ZJ\_G0601 filter resulting in a spectral resolution of $\approx1.7\,\AA$, as obtained from the measurement of 
the full width at half maximum (FWHM) of arc lamp lines. The K$_{\rm l}$-band observations were done using the Kl\_G5607 grating and HK\_G0603 filter and resulted 
 in a spectral resolution of FWHM$\approx3.3\,\AA$.

The total exposure time at each band was 4800\,s, consisting of 8 individual on-source exposures. A small spatial dithering of 0\farcs1 
was applied between on-source exposures in order to correct the data from bad pixels.The FWHM of the spatial profile of the star
 was 0\farcs13$\pm$0\farcs02 for the J-band and 0\farcs15$\pm$0\farcs03 for the K$_{\rm l}$-band.

The data reduction was accomplished using tasks contained in the {\sc nifs}
package which is part of {\sc gemini iraf} package as well as generic {\sc
iraf} tasks. The data reduction followed the standard procedure, which includes  
the  trimming of the images, flat-fielding, cosmic ray rejection, sky subtraction, wavelength and s-distortion calibrations.
In order to remove telluric absorptions from the galaxy spectra we observed the standard star HIP\,15925
 just after the J-band observations of the galaxy and  HIP\,10559 just before the K$_{\rm l}$-band observations.
The galaxy spectra was divided by the normalized spectrum of the tellurc standard star using the {\sc nftelluric} task of 
the {\sc nifs.gemini.iraf} package. The galaxy spectra were flux calibrated by interpolating a blackbody function 
to the spectrum of the telluric standard and the J and  K$_{\rm l}$-bands data cubes were constructed with an 
angular sampling of 0\farcs05$\times$0\farcs05 for each individual exposure. Finally the individual data cubes 
were combined using a sigma clipping algorithm in order to eliminate bad pixels and remainding cosmic rays by mosaicing the dithered spatial positions.
 The final J and K$_{\rm l}$ data cubes contain about 4200 individual spectra and cover the central 3$^{\prime\prime}\times$3$^{\prime\prime}$, corresponding to 
$\approx$700$\times$700 square parsecs at the galaxy.
\subsection{The 3.6\,cm radio image}

The 3.6~cm Very Large Array
  (VLA) radio image was made using data taken on 31 December
1992, initially published in \citet{nagar99}, and reprocessed
as described in \citet{mundell09}.

\section{Results}

In the top-left panel of Fig.\,\ref{espectro} we present a large scale $R$-band image of Mrk\,1066 from the Palomar
Observatory Sky Survey \citep{lasker96}. In 
the top-right panel we present an image obtained from the NIFS data cube for the  continuum around 2.22$\,\mu$m. In the bottom panels 
we present two characteristic IFU spectra: the nuclear one and a spectrum 
from a location at 0\farcs5 north-west of the nucleus. % (position A). Both spectra correspond to an aperture of  0$\farcs$25$\times$0$\farcs$25.
In the J-band we identified the [P {\sc ii}]~emission lines at 1.14713 and 1.18861\,$\mu$m, the [Fe\,{\sc ii}] emission lines at 
1.25702, 1.27912, 1.29462, 1.29812, 1.32092 and 1.32814\,$\mu$m, the H\,{\sc i}~Pa$\beta\,\lambda$1.28216\,$\mu$m, the He\,{\sc ii} line at 1.16296\,$\mu$m
 and the [S\,{\sc ix}] coronal line at 1.25235\,$\mu$m at the nucleus. The H$_2$ emission lines at 2.12183, 2.15420, 2.22344, 2.24776, 2.40847, 2.41367, 2.42180, 
2.43697 and 2.45485\,$\mu$m are identified in the K-band spectra, as well as the  H\,{\sc i}~Br$\gamma\,\lambda$2.16612\,$\mu$m, the  
He\,{\sc i}\,$\lambda$2.14999\,$\mu$m and the [Ca\,{\sc viii}]\,$\lambda$2.32204\,$\mu$m coronal line at the nucleus. In the K-band spectra
 we have also identified the CO stellar absorption band heads around 2.3\,$\mu$m.

\subsection{The continuum emission}\label{sec_continuum}

Figure\,\ref{cnt} presents optical and near-IR continuum images of the central 3$^{\prime\prime}\times$3$^{\prime\prime}$ of Mrk\,1066. 
The top left panel shows an optical image obtained with the HST 
 Wide Field Planetary Camera 2 (WFPC2) through the filter F606W \citep{malkan98}. This image %is shown in 
%arbitrary flux units and 
presents two spiral arms oriented approximately along  PA=128$^\circ$ 
(approximately coincident with the major axis of the galaxy), which extend up to 1\farcs2  to both sides of the nucleus. 
Along the spiral arms there are several knots with enhanced emission. This image is in good agreement with the optical continuum images 
presented by \citet{bower95}.
% To north-east of the nucleus there is a low intensity structure resembling a dusty spiral arm,
% with extension of about 2$^{\prime\prime}$  and ends at about 1$^{\prime\prime}$\,east of the nucleus. To south-east of the nucleus a similar 
% structure can be identified. 

\begin{figure}
\centering
\includegraphics[scale=0.6]{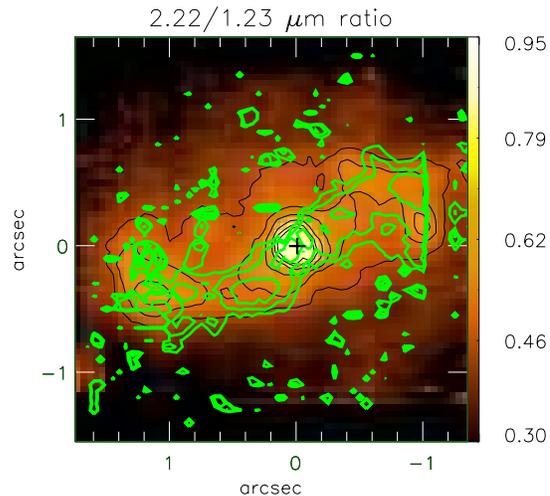}
\caption{K/J continuum ratio obtained by averaging continuum regions (in units of erg\,s$^{-1}$\,cm$^{-2}$\, \AA$^{-1}$)
 centered on 2.22\,$\mu$m and 1.23\,$\mu$m. 
The thick contours are from the reddening map for Fig.~\ref{ratio}.}
\label{ratio_cnt}
\end{figure}

In the top right panel of Fig.\,\ref{cnt} we present an HST H-band image obtained  
with the Near Infrared Camera and Multi-Object Spectrometer (NICMOS) through the filter F160W (proposal 
ID 7330 by J. S. Mulchaey).% in arbitrary flux units.
 In the bottom panels of Fig.\,\ref{cnt} 
we present the J and K$_{\rm l}$-band continuum images obtained from the NIFS data cubes 
by averaging continuum regions centered at 1.23$\,\mu$m and  2.22$\,\mu$m.%, respectively, in logarithmic flux units [log(erg\,s$^{-1}$\,cm$^{-2}$)]. 
%The thick contours overlaid on the K-band map are from the HST optical image. 
The near-IR continuum is more extended along PA=128$^\circ$, which is similar to the orientation of the spiral arms observed in the optical continuum and approximately the 
orientation of the major axis of the galaxy.  The spiral arms can also be seen in the near-IR, although with ``less contrast'' than in the optical, 
probably due to the stronger effect of the dust and contribution of line emission in the optical.
 A comparison between the J and K continuum shows that the extra-nuclear emission is bluer than the nuclear one, indicating the presence 
of dust nucleus. This behavior can be observed in Fig.~\ref{ratio_cnt}, which presents a K/J continuum ratio map obtained 
from the NIFS data cubes by averaging the flux within spectral window of 50~\AA\ in regions free of emission and absorption lines centered at 2.22\,$\mu$m 
 and  1.23$\mu$m.

\subsection{Emission-line flux distributions}

As shown in Fig.~\ref{espectro}, several emission lines can be observed in the spectra. 
Table\,\ref{fluxes} presents  the measured fluxes for the  emission lines within an aperture of 0\farcs25$\times$0\farcs25 
at the four positions identified in the top-right panel of Fig.~\ref{espectro}: the nucleus, position\,A at 0\farcs5\,north-west 
from the nucleus, position\,B at 0\farcs5\,south-east and position\,C at 0\farcs5\,north-east from the nucleus. 
 Most emission lines peak at the nucleus,
except for [Fe\,{\sc ii}] and [P\,{\sc ii}] which peak at 0\farcs5\,north-west of the nucleus.

\begin{table*}
\centering
\caption{Measured emission-line fluxes for the four positions marked in Fig.~\ref{espectro} within 0\farcs25$\times$0\farcs25 aperture in 
units of 10$^{-16}$\,erg\,s$^{-1}$\,cm$^{-2}$.}
\vspace{0.3cm}
\begin{tabular}{l l c c c c}
\hline
$\lambda_{vac} {\rm(\AA)}$    & ID                    & Nucleus         & A               &  B              & C     \\
\hline
11471.3  & [P {\sc ii}]\,$^1D_3-^3P_1$                & 5.07$\pm$2.23   & 2.75$\pm$0.71    &  1.50$\pm$0.47 &   -- \\
11629.6  & He\,{\sc ii}\,$7-5$                        & 2.41$\pm$1.52   &  0.99$\pm$0.32   &  0.67$\pm$0.13 &   -- \\
11886.1  & [P {\sc ii}]\,$^1D_2-^3P_2$                & 6.21$\pm$0.69   &  6.54$\pm$0.87   &  2.80$\pm$0.38 & 	0.41$\pm$0.28 \\
12523.5  & [S\,{\sc ix}]\,$^3P_1-^3P_2$               & 0.60$\pm$0.37   &   --             &   --           &	--	       \\
12570.2  & [Fe\,{\sc ii}]\,$a^4D_{7/2}-a^6D_{9/2}$    & 18.88$\pm$0.80  &  21.15$\pm$1.82  &  13.06$\pm$0.95&	5.76$\pm$1.21 \\
12706.9  & [Fe\,{\sc ii}]\,$a^4D_{1/2}-a^6D_{1/2}$    & 0.52$\pm$0.41   &  0.96$\pm$0.38   &  0.43$\pm$0.17 &	0.14$\pm$0.11	\\
12712.1  & [Fe\,{\sc iv}]\,$^2G^5_{7/2}-^2D^5_{5/2}$ ?& 0.21$\pm$0.19   &   --             &    --          &	--	        \\
12791.2  & [Fe\,{\sc ii}]\,$a^4D_{3/2}-a^6D_{3/2}$    & 0.94$\pm$0.35   &  1.34$\pm$0.65   &  1.27$\pm$0.32 &	0.36$\pm$0.25	\\
12821.6  &  H\,{\sc i}\,Pa$\beta$                     & 41.18$\pm$4.26  &  25.17$\pm$1.98  &  26.41$\pm$1.98&	6.79$\pm$0.55  \\
12946.2  & [Fe\,{\sc ii}]\,$a^4D_{5/2}-a^6D_{5/2}$    & 1.91$\pm$0.87   &  1.71$\pm$0.16   &  0.82$\pm$0.29 &	0.70$\pm$0.13	\\
12981.2  & [Fe\,{\sc ii}]\,$a^4D_{3/2}-a^6D_{1/2}$    & 1.72$\pm$0.55   &   0.53$\pm$0.22  &  0.94$\pm$0.23 &	0.46$\pm$0.29	 \\
13209.2  & [Fe\,{\sc ii}]\,$a^4D_{7/2}-a^6D_{7/2}$    & 4.99$\pm$0.73   &  6.32$\pm$0.64   &  4.03$\pm$0.52 &	2.21$\pm$0.26   \\
13281.4  & [Fe\,{\sc ii}]\,$a^4D_{5/2}-a^6D_{3/2}$    & 1.54$\pm$1.05   &  1.31$\pm$0.59   &  0.54$\pm$0.40 &	--      	 \\

21218.3  & H$_2$2\, 1-0S(1) 	  		      & 13.41$\pm$1.02  & 11.74$\pm$0.49    & 5.27$\pm$0.48  &	 2.52$\pm$0.11	\\
21499.9  &  He\,{\sc i}\,$^3S_1-^3P^0_1$              & 0.40$\pm$0.21   & 0.11$\pm$0.21    &  0.27$\pm$0.19 &	 0.18$\pm$0.07	\\
21542.0  &  H$_2$\,1-0\,S(2)  	                      & 0.51$\pm$0.23   & 0.45$\pm$0.50    & 0.25 $\pm$0.18 &	 0.16$\pm$0.04	\\
21661.2  & H\,{\sc i}\,Br$\gamma$                     &17.14$\pm$1.16   & 10.59$\pm$0.60    & 9.81$\pm$0.53  &	 1.28$\pm$0.08	\\
22234.4  &  H$_2$\,1-0\,S(0)  	                      & 2.99$\pm$0.92   & 2.63$\pm$0.64    & 1.96$\pm$0.72  &	 0.65$\pm$0.06	\\
22477.6  &  H$_2$\,2-1\,S(1)  	                      & 1.47$\pm$0.62   & 1.17$\pm$0.38    & 0.67$\pm$0.33  &	 0.33$\pm$0.02	\\
23220.4  &[Ca\,{\sc viii}]\,$^2P^0_{3/2}-^2P^0_{1/2}$ &4.64$\pm$1.42    &  --              &   --           & 	 --	\\
24084.7  &  H$_2$\,1-0\,Q(1)   	                      &10.04$\pm$1.33   & 8.14$\pm$0.45    & 4.82$\pm$0.86  &	  1.42$\pm$0.11 \\
24136.7  &  H$_2$\,1-0\,Q(2)   	                      & 3.03$\pm$0.47   & 2.47$\pm$0.42    & 1.47$\pm$0.41  &	 0.49$\pm$0.10	\\
24218.0  &  H$_2$\,1-0\,Q(3)   	                      & 10.61$\pm$1.05  & 8.32$\pm$0.57    & 5.23$\pm$0.56  &	  1.50$\pm$0.21 \\
24369.7  &  H$_2$\,1-0\,Q(4)   	                      & 2.73$\pm$1.45   & 2.26$\pm$0.38    & 1.30$\pm$0.48  &	 0.47$\pm$0.09 \\
24548.5  & H$_2$\,1-0\,Q(5)   	                      & 7.08$\pm$2.18   & 5.41$\pm$0.43    & 2.82$\pm$0.77  &	 4.36$\pm$2.24 \\  
\hline

\end{tabular}
\label{fluxes}
\end{table*}

\begin{figure*}
\centering
\includegraphics[scale=1.1]{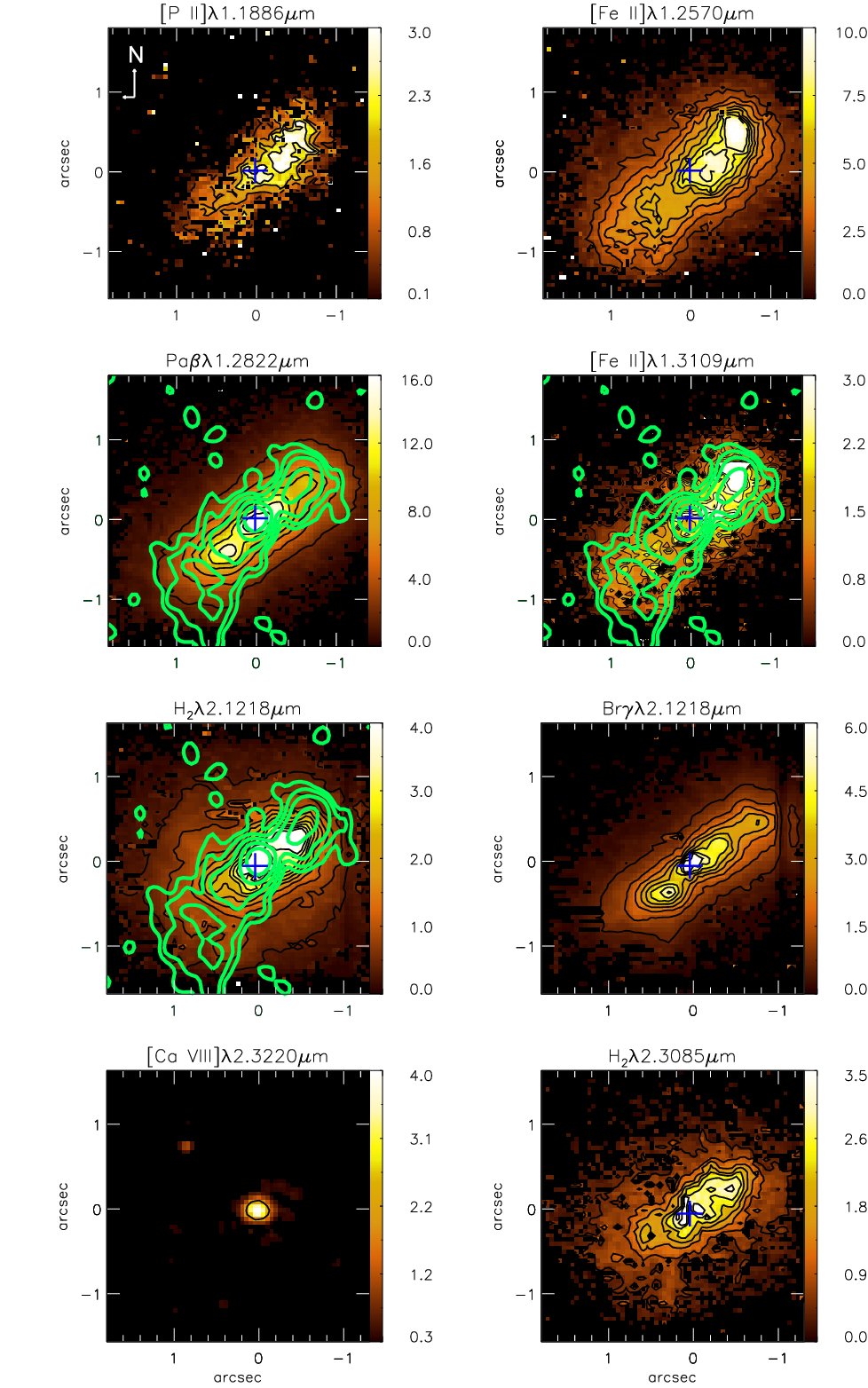}
\caption{Integrated flux maps for the [P\,{\sc ii}]$\lambda$1.8861$\,\mu$m, [Fe\,{\sc ii}]$\lambda$1.2570$\,\mu$m, 
Pa$\beta$, [Fe\,{\sc ii}]$\lambda$1.3109$\,\mu$m, H$_2\lambda$2.1218$\,\mu$m, Br$\gamma$,  [Ca\,{\sc viii}]$\lambda$2.3220$\,\mu$m and  
H$_2\lambda$2.1218$\,\mu$m emission lines in units of 10$^{-17}$\,erg\,s$^{-1}$\,cm$^{-2}$. The 
central cross marks the position of the continuum peak. The thick contours overlaid on the Pa$\beta$, [Fe\,{\sc ii}]$\lambda$1.2570$\,\mu$m 
and H$_2\lambda$2.1218$\,\mu$m maps are from the 3.6\,cm radio continuum image.}% from \citet{nagar99}.}
\label{flux}
\end{figure*}

In order to map the gas emission we have integrated the flux under the emission-line profiles and subtracted the underlying continuum obtained  
from two spectral windows, one from each side of the profile. Figure\,\ref{flux} presents the resulting flux maps.
% The thick contours are from the Very Large Array (VLA) 3.6\,cm radio continuum image
% from \citet{nagar99}, reprocessed by Nagar in 2004 (private counication) 
%and the central crosses mark the position of the nucleus defined as the near-IR 
%continuum peak.% Fluxes are given in units of 10$^{-17}$\,erg\,s$^{-1}$\,cm$^{-2}$.
%Fig.\,\ref{flux} shows that, 
For most emission lines, the highest intensity levels  most extended 
along  PA$\approx135/315^\circ$, which is the orientation of the radio structure.  A detailed inspection of each panel
 reveals distinct structures for each ionization level. The \pii, \feii\  and \h2\ emission lines are $\sim$2 times brighter to the north-west 
than to the south-east, while \pb\ and \br\ are approximately equally bright to both sides of the nucleus. 
The peak fluxes of the [P\,{\sc ii}] and [Fe\,{\sc ii}] emission lines occur at 
$\approx$0\farcs85\,north-west of the nucleus, approximately at the same position where there is a hot spot in the radio map. 
The H\,{\sc i} recombination-line emission peaks at the nucleus and presents a secondary peak at $\approx$0\farcs5\,south-east from it.
Another difference between the H\,{\sc i} and [Fe\,{\sc ii}] emission lines is that the flux distribution of the latter traces the radio structure 
better than the former. The H$_2$ emission also presents a peak at the nucleus and a secondary peak at
 $\approx$0\farcs65\,north-west of the nucleus, in a region between the nucleus and the 
radio hot spot. The emission in the coronal lines of [Ca\,{\sc viii}]$\lambda$2.3220$\,\mu$m and [S\,{\sc ix}]\,$\lambda1.2524\,\mu$m are unresolved by our observations.
 %The  [S\,{\sc ix}] line was not detected in previous lower spatial resolution observations \citep{knop01,rodriguez-ardila05a}

\subsection{Line-ratio maps}\label{sec_ratio}

\begin{figure*}
\centering
\includegraphics[scale=0.9]{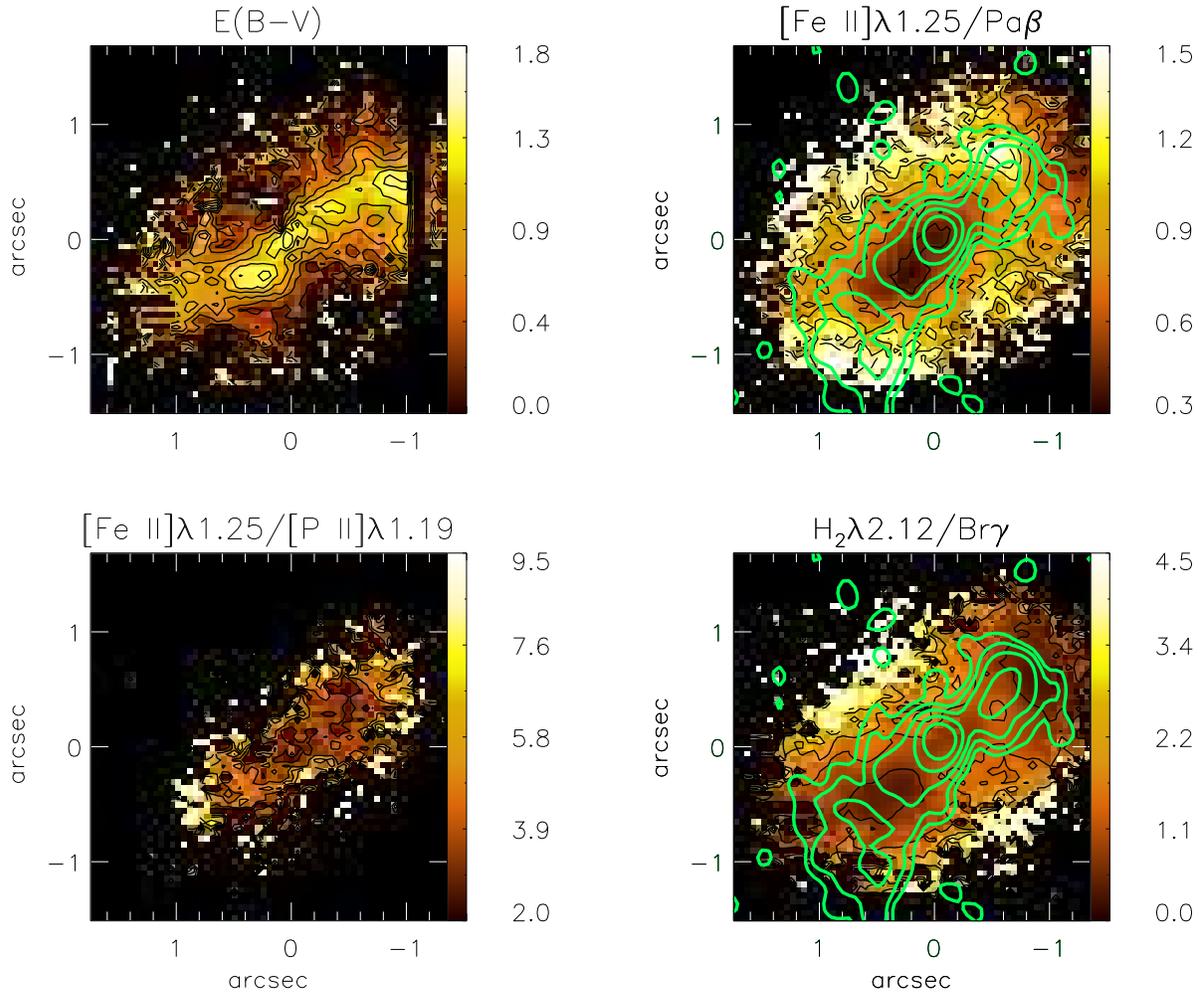}
\caption{Top: Reddening map obtained from the Pa$\beta$/Br$\gamma$ line ratio (left) and 
[Fe\,{\sc ii}]$\lambda$1.2570$\,\mu$m/Pa$\beta$ line ratio map (right). Bottom: 
[Fe\,{\sc ii}]$\lambda$1.2570$\,\mu$m/[P\,{\sc ii}]$\lambda$1.8861$\,\mu$m line-ratio
 map (left) and H$_2\lambda$2.1218$\,\mu$m/Br$\gamma$ ratio (right). The thick contours
 overlaid to the \feii/\pb\ and \h2/\br\ maps are from the radio image.}
\label{ratio}
\end{figure*}

In order to investigate the excitation mechanisms of the [Fe\,{\sc ii}] and H$_2$ emission lines we constructed flux ratio maps, 
 shown in Figure\,\ref{ratio}. In this figure we also present a reddening map obtained from the Pa$\beta$/Br$\gamma$ line ratio as 
\begin{equation}
 E(B-V)=4.74\,{\rm log}\left(\frac{5.88}{F_{Pa\beta}/F_{Br\gamma}}\right),
\end{equation}
where $F_{Pa\beta}$ and $F_{Br\gamma}$ are the fluxes of $Pa\beta$ and $Br\gamma$ emission lines, respectively. We have used the reddening 
law of \citet{cardelli89} and adopted the intrinsic ratio $F_{Pa\beta}/F_{Br\gamma}=5.88$ corresponding to case B recombination \citep{osterbrock06}. 
The resulting $E(B-V)$ map is shown in the top left panel of Fig.\,\ref{ratio}. The highest values are observed in a S-shaped structure, 
resembling the spiral arms observed in the continuum image overlaid as  thick contours on the reddening map.
%[resembling an spiral arm, oriented along the PA$\approx$128$^\circ$]. 
Typical values within this structure are $E(B-V)\approx1.0$, 
with several knots of higher values reaching $E(B-V)\approx1.7$. In regions away from the center of the S-shaped structure the 
reddening decreases to $E(B-V)\approx0.4$ or bellow.  The errors in $E(B-V)$ are approximately 0.15. 

A comparison between the emission-line reddening and the continuum ratio map is shown in Fig.~\ref{ratio_cnt}. The thick contours, representing 
the reddening map, show that the S-shaped structure is also present in the continuum ratio map as a region of higher K/J ratio. In order to estimate the 
reddening in the continuum, we note that, from Fig.~\ref{ratio_cnt}, a continuum ratio (K/J)$_1\approx$0.4 corresponds to the locations with $E(B-V)_1\approx0.4$
 in the gas. If the higher (K/J)$_2$ ratio along the S-shaped structure is due to reddening, the corresponding $E(B-V)_2$ can be calculated from 
\begin{equation}
 \chi[E(B-V)_1-E(B-V)_2]={\rm log}\left[\frac{{\rm (K/J)_1}}{{\rm(K/J)_2}}\right],
\end{equation}
where $\chi=-0.4(A_K/A_V-A_J/A_V)R_V=0.2083$, where  $(A_K/A_V)=0.114$, $(A_J/A_V)=0.282$ and $R_V=3.1$ for the extinction law of \citet{cardelli89}. Using 
${\rm(K/J)_2}=0.5$ measured in the S-shaped structure (from Fig.~\ref{ratio_cnt}) we obtain  reddening for the continuum of $E(B-V)_2\approx0.9$, which
 is somewhat smaller than the one observed for the emitting gas (top-left panel of Fig.~\ref{ratio}). Nevertheless, 
right at the nucleus the K/J ratio is much higher than in the rest of the S-shaped structure, what indicates that it is not due to reddening. The nature 
of this redder continuum will be discussed in Sec.~\ref{disc_dust}.

In the top-right panel of Fig.\,\ref{ratio} we present the [Fe\,{\sc ii}]$\lambda$1.2570$\,\mu$m/Pa$\beta$ intensity ratio, which can be used to 
investigate the excitation mechanism of the [Fe\,{\sc ii}] emission. Seyfert galaxies present typical values for this ratio between 0.6 and 2.0.
\citep[e.g.][]{riffel06,rodriguez-ardila04,rodriguez-ardila05a,storchi-bergmann09}. The smallest value observed in Mrk\,1066 is
[Fe\,{\sc ii}]$\lambda$1.2570$\,\mu$m/Pa$\beta=0.33\pm0.03$ at 0\farcs3\,south-east from the nucleus, while the highest values are 
observed in regions located at distances larger than 1$^{\prime\prime}$, reaching values of up to 1.5.

Another line ratio that can be used to investigate the [Fe\,{\sc ii}] excitation mechanism is 
[Fe\,{\sc ii}]$\lambda$1.2570$\,\mu$m/[P\,{\sc ii}]$\lambda$1.8861$\,\mu$m \citep[e.g.][]{oliva01,storchi-bergmann09}. We present this ratio 
in the bottom-left panel of Fig.\,\ref{ratio}. The lowest values of $\approx3$ are observed at the nucleus and in regions to north-west at small distances from the 
nucleus, while the highest values up to 9.5 are observed at larger distances from the nucleus.

In the bottom-right panel of Fig.\,\ref{ratio} we present  the H$_2\lambda$2.1218$\,\mu$m/Br$\gamma$ ratio map, which is useful to investigate 
the excitation of the H$_2$ emitting gas. This ratio presents 
values ranging from 0.6 to 2 for Seyfert galaxies \citep[e.g.][]{riffel06,rodriguez-ardila04,rodriguez-ardila05a,storchi-bergmann09}. In Mrk\, 1066, 
the lowest values of $0.29\pm0.06$ are observed at 1$^{\prime\prime}$\,north-west from the nucleus and the highest 
values of up to 2.6 are observed at distances of $>$0\farcs7 in regions away from the radio structure. Another region of small values ($\approx0.4$) 
 is observed at $\approx$0\farcs5\,south-east of the nucleus, approximately coincident with the secondary peak observed in the 
H\,{\sc i} recombination-lines flux maps (see Fig.\,\ref{flux}), while at the nucleus H$_2\lambda$2.1218$\,\mu$m/Br$\gamma=0.75\pm0.07$.

\section{Discussion}

\subsection{Dust emission} \label{disc_dust}

As observed in Fig.\,\ref{ratio_cnt} (see also Figs.~\ref{espectro} and \ref{cnt}) the extra-nuclear continuum  is bluer than the 
nuclear continuum. At the nucleus the ratio between the continuum at 2.22\,$\mu$m and at 1.23\,$\mu$m is $\approx 1$, while at 
larger distances from the nucleus, typical values are  $\approx 0.45$ along  PA$\approx128^\circ$ and even smaller at locations away from this PA
 ($\lesssim0.38$).  In order to to investigate the nature of the K-band excess at the nucleus, 
we have extracted a nuclear spectrum within an aperture of 0\farcs25 radius, which includes the nuclear flux down to 20\% of the peak continuum 
flux in the K band. An extra-nuclear spectrum  for a circular ring within  radii 0\farcs25$\leq r \leq$0\farcs35 was extracted
 to represent the  stellar population. This spectrum was normalized according to the ratio of the nuclear/extra-nuclear extraction apertures. Both
 spectra were corrected for extinction using the reddening values obtained in Sec.~\ref{sec_ratio}:  $E(B-V)\approx1.03$ for the nucleus and 
 $E(B-V)\approx0.69$ for the stellar population.

% In the top panels of Figure\,\ref{fit}
% we present the resulting spectra.  The nuclear K-band spectrum is nearly flat, while the extra-nuclear falls to higher wavelengths. 

Under the assumption that the underlying stellar population in the nuclear spectrum is the same as in the extra-nuclear spectrum, we have  
subtracted the latter from the former. The resulting nuclear continuum was then modeled by the sum of two components: a blackbody function, which dominates the emission in the K band 
and a power-law, which is important for the J-band emission. The fit was performed using the {\it nfit1d} task of the {\sc stsdas iraf} 
package selecting only spectral regions not affected by line emission. The best fit was obtained for a power-law $F_\lambda \propto \lambda^{-2.4\pm0.3}$ 
and a blackbody function with temperature $T=863\pm30$\,K. 

%In order to evaluate the effect of the reddening in the continuum emission, we used the reddening obtained in Sec.~\ref{sec_ratio} 
%for the NLR emitting, to correct the nuclear and extra-nuclear spectra. The reddening obtained for the nuclear aperture is $E(B-V)\approx1.03$ and for extra-nuclear
% spectrum  we obtained $E(B-V)\approx0.69$. Using these values to correct the 
%nuclear and extra-nuclear spectra and fitting again the subtracted spectrum, we obtain spectral index and dust temperature similar to 
%the ones obtained without reddening correction ($F_\lambda \propto \lambda^{-1.02\pm0.15}$ and $T=807\pm30$\,K), suggesting that the reddening 
%effect is similar in both, nuclear and extra-nuclear, spectra.  

\begin{figure*}
%\centering
\includegraphics[scale=1.1]{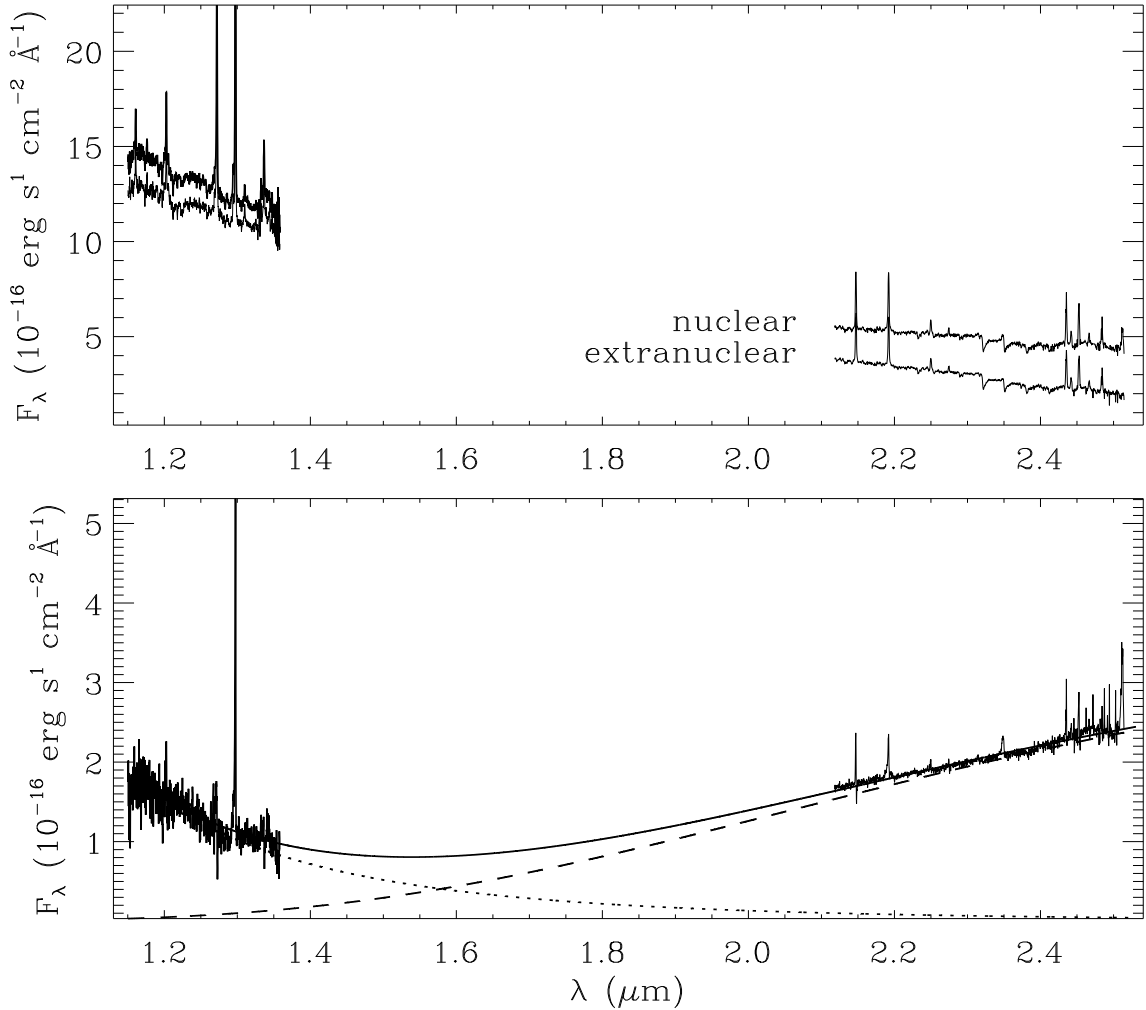}
\caption{Top: Nuclear and extra-nuclear spectra of Mrk\,1066. The nuclear spectrum was extracted within a circular aperture with radius 
0\farcs25, while the extra-nuclear spectrum was integrated within an aperture with inner radius 0\farcs25 and outer radius of 0\farcs35. 
Both spectra were corrected by reddening. The extra-nuclear spectrum was multiplied by 1.45, after normalization to the same aperture of the nuclear one, 
in order to eliminate the stellar absorption in the subtracted spectrum. Bottom: Difference between the nuclear 
and extra-nuclear spectra, together with a fit (solid line) considering the contribution of a power-law (dotted line) plus a 
blackbody function (dashed line).}
\label{fit}
\end{figure*}

We then considered the possibility that the contribution of the stellar population to the nuclear spectrum is higher than the one derived only from the 
normalization of the fluxes to the same extraction aperture. This possibility is suggested 
by the fact that the stellar CO absorption bands in the nuclear spectrum do not disappear after the subtraction of the extra-nuclear spectrum. In order 
to completely eliminate the stellar absorptions the extra-nuclear spectrum had to be multiplied by a factor of 1.45. After this correction 
we obtain $F_\lambda \propto \lambda^{-4.7\pm0.4}$ and $T=829\pm30$\,K. In the top panel of Fig.~\ref{fit} we show the nuclear and extra-nuclear 
spectra corrected by extinction with the latter properly normalized as described above. The bottom panel shows the subtracted spectrum, as well as the resulting
 fit as a continuous line. The blackbody function is shown as a dashed line and the power-law as a dotted line.

%If the contribution of the underlying stellar population is higher than the one evaluated from the 0\farcs25$\leq r \leq$0\farcs35 
%extra-nuclear spectrum it could modify the parameters derived by fitting. A comparisson of the nuclear and extra-nuclear J band spectra 
%shows that the stellar population could contributes at maximum 10\% more than the contribution obtained from the  extra-nuclear 
%spectrum, since if we multiply the extra-nuclear spectrum by a factor higher than 1.1 it would be stronger than the J band nuclear spectrum. 
%Doing this correction, we obtain similar parameters than the ones derived above, within the uncertainties.  

Recently, \citet{rogerio09}\footnote{This is not the present author, but his brother.} presented stellar population synthesis for a sample of Seyfert galaxies, including Mrk\,1066, 
using  near-IR spectra extracted within an aperture of 1\farcs6$\times$0\farcs8. Besides 
synthetic Simple Stellar Population (SSP) models, the authors considered the contribution of a power-law to represent the featureless 
continuum from the AGN and five blackbody
 functions for temperatures ranging from 800\,K to 1400\,K in order to account for possible dust emission.  They concluded that the 
dust emission is not necessary in the case of Mrk\,1066.  Nevertheless, as shown in Fig.\,\ref{fit}, a K-band excess is
 clearly present in our nuclear spectrum. \citet{rogerio09} did not identified this component 
probably due to their larger aperture, which covers an area more than 6 times ours. Unresolved dust emission from the nucleus of Seyfert galaxies 
 is commonly observed and is usually attributed to the dusty torus postulated by the Unified Model 
\citep[e.g.][]{riffel09a,riffel09b,rogerio09,rodriguez-ardila06,rodriguez-ardila05b}.   Previous works have shown that 
the dust emission in the near-IR for Seyfert\,1 galaxies give blackbody temperatures $T\gtrsim1200\,$K,
 while for Seyfert\,2 galaxies the blackbody temperature is usually $T\lesssim900\,$K, in agreement with our result for Mrk\,1066. 
 This difference can be interpreted as due to the fact that in Seyfert~1 galaxies one can see the inner wall of the torus, where the temperatures 
are close to the evaporation temperature of graphite and silicate grains \citep{barvainis87,granato94}. In Seyfert~2 galaxies we 
do not see the inner wall, only dust located at larger distances from the nucleus, heated by a more diluted radiation field, 
part of which is also blocked by clouds in the inner region of the torus \citep[e.g.][]{elitzur08}. Since the K-band continuum of Mrk\,1066 is 
unresolved by our observations, we can constraint the location of the emitting dust to within the inner $\approx18~$pc, although, as 
discussed in \citet{riffel09b}, the dust is probably located much closer in.

We can derive the mass of the emitting dust from the parameters obtained from the fit of the blackbody function, together with assumptions about the 
physical properties of the dust grains.  We estimate the infrared  luminosity of each  grain assuming that the dust composition is
 graphite, by
\begin{equation}
 L^{\rm gr}_{\rm \nu,ir} = 4\pi^2a^2 Q_\nu B_\nu(T_{\rm gr})~~~~[{\rm erg\,s^{-1}\,Hz^{-1}}]
\label{lgr}
\end{equation}
where $a$ is the grain radius, $Q_\nu=q_{ir}\nu^\gamma$ is the absorption efficiency and $B_\nu(T_{\rm gr})$ 
is its spectral energy distribution assumed to be a Planck function with temperature $T_{\rm gr}$ \citep{barvainis87}.  The ratio  between the total 
luminosity of the hot dust ($L_{\rm ir}^{\rm HD}$) and that of one  grain  ($L_{\rm ir}^{\rm gr}$) gives 
the number of dust grains $N_{\rm HD}= \frac{L_{\rm ir}^{\rm HD}}{L_{\rm ir}^{\rm gr}}$. $L_{\rm ir}^{\rm HD}$ was obtained by integrating the 
flux under the Planck function fitted to the nuclear spectrum and $L_{\rm ir}^{\rm gr}$ was obtained by the integration of 
equation\,\ref{lgr} for a temperature of $T_{gr}=829\,$K assuming $a=0.05\,\mu$m, 
$q_{ir}=1.4\times10^{-24}$ and $\gamma=1.6$ \citep{barvainis87,kishimoto07}. Finally, the mass of the emitting hot dust is obtained as:

\begin{equation} 
M_{\rm HD}\approx\frac{4\pi}{3}a^3 N_{\rm HD}\rho_{\rm gr}.
\end{equation}

Assuming a graphite density of $\rho_{\rm gr}=2.26\,$g\,cm$^{-3}$ \citep{granato94}, we obtain
 $M_{\rm HD}=(1.4\pm\,0.5)\times10^{-2}\,{\rm M_\odot}$. The derived mass is comparable to values obtained for other Seyfert galaxies, 
which are in the range $10^{-6}-10^{-2}\,{\rm M_\odot}$ \citep[e.g.][]{riffel09b,rogerio09}.

\subsection{Flux distributions}\label{disc_distr}

The flux distributions shown in Fig.\,\ref{flux}, are  more more extended along the 
PA$\approx$135/315$^\circ$ and resemble those in optical lines \citep{bower95,stoklasova09}. Using HST 
imaging, \citet{bower95} observed a narrow ``jetlike'' feature in the [O\,{\sc iii}]+H$\beta$ emission, extending to 1\farcs4~north-west 
from the nucleus along PA=315$^\circ$. They concluded that this image is dominated by [O\,{\sc iii}] emission and thus that the high 
excitation gas is strongly concentrated in the jet. To the south-east the [O\,{\sc iii}] emission is much fainter than to the north-west, 
similarly to what we observe in the \feii~and \pii~emission. \citet{bower95} present also a \ha+\nii~narrow-band image, which shows the same ``jetlike'' 
structure, but with a flux distribution broader then in \oiii, with emission to both sides of the nucleus, similar to our \br\ and \pb\ flux distributions. 
 The 3.6\,cm radio map presents a bipolar jet at approximately 
 the same orientation of the emission-line flux distributions 
(see Fig.\,\ref{flux}). By comparing the radio and optical line emission, \citet{bower95} suggest that the low- and high-excitation gas present 
distinct spatial distributions, with the former preferentially located in the galaxy plane and the latter in the jet region. 
They explain the lower intensity of \oiii~to south-west of the nucleus as being due to obscuration of the jet by the galactic plane.

\begin{figure}
\centering
\includegraphics[scale=0.47]{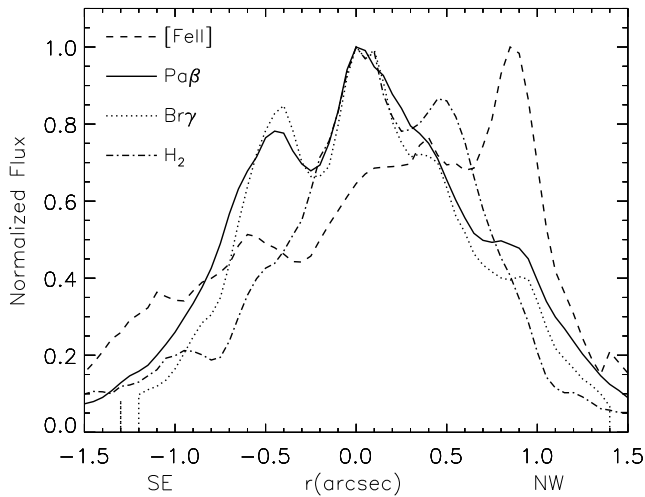}
\caption{One-dimensional cuts from the flux distributions in \feii, \pb, \br~and \h2$\lambda2.12\mu$m emission-lines extracted
 of a pseudo-slit with 0\farcs25 width oriented along  PA=135$^\circ$. The fluxes are normalized to the peak value. }
\label{oned}
\end{figure}

A detailed analysis of the line flux distributions reveal some  differences between lines from distinct ionization levels. In order
to do this comparison, we extracted  one-dimensional cuts along  PA=135$^\circ$ for a pseudo-slit with 0\farcs25 width. The resulting 
one-dimensional flux distributions for the \feii, \pb, \br~and \h2$\lambda2.12\mu$m emission lines are shown in Figure\,\ref{oned}. 
As observed in this figure (see also the two-dimensional maps from Fig.\,\ref{flux}), the \feii\ flux distribution peaks at $\approx$0\farcs85 
north-west of the nucleus, approximately 
coincident with a hot spot observed in the radio map, and is fainter to the south-east, in good agreement with the \oiii~image.
 The H recombination lines and \h2 line emission peak at the nucleus. The  H lines have a
 secondary peak at 0\farcs5 
south-east of the nucleus, which is not observed in the \h2~and \feii~flux distributions. The \h2 emission presents a secondary peak 
to north-west, between the nucleus and the radio hot spot. \citet{knop01} obtained J and K long-slit spectroscopy of Mrk\,1066 with a seeing of 
0\farcs7. They present flux measurements for near-IR emission lines along  PA=45$^\circ$ and 
  PA=135$^\circ$, which can be directly compared with our one-dimensional cuts and flux maps (Figs.\,\ref{oned} and \ref{flux}). 
In general, our flux distributions are in good agreement with theirs, although much  more details are revealed by our measurements. 
In particular, the secondary peaks observed in the H emission lines and the \h2~nuclear peak are not present in the flux distributions 
presented by \citet{knop01}. We attribute these differences to our higher spatial resolution.
 
Our \feii~and H recombination lines present similar flux distributions to those of the \oiii~and \ha+\nii~images of \citet{bower95}, 
respectively. We conclude that the \feii~emitting gas extends to high galactic latitudes and is approximately co-spatial with the radio jet, while 
the \br~and \pb~emission originate mostly from the galactic plane. The \h2~flux is more uniformly distributed over the whole IFU field, suggesting also
 an important contribution from gas in the galactic plane, although part of the \h2\ emitting gas close to the nucleus may be also outflowing to high latitudes,
 as judged from the highest flux to the north-west than to the south-east (Fig.~\ref{flux}). 
These conclusions are supported by the observation of distinct 
kinematics for these emission lines, with the \feii\ originating in a much more disturbed gas than the H and \h2 emitting (Riffel et al., in preparation).
 This result is also supported by near-IR IFU observations
 of other Seyfert galaxies, which show that the \h2~and ionized gas have distinct distributions and kinematics, with the former usually restricted 
to the galactic plane and the latter extending to higher latitudes 
\citep[e.g.][]{riffel06,riffel08,riffel09a,storchi-bergmann09}. 

\subsection{Diagnostic diagram} \label{disc_diagn}

\begin{figure}
\centering
\includegraphics[scale=0.92]{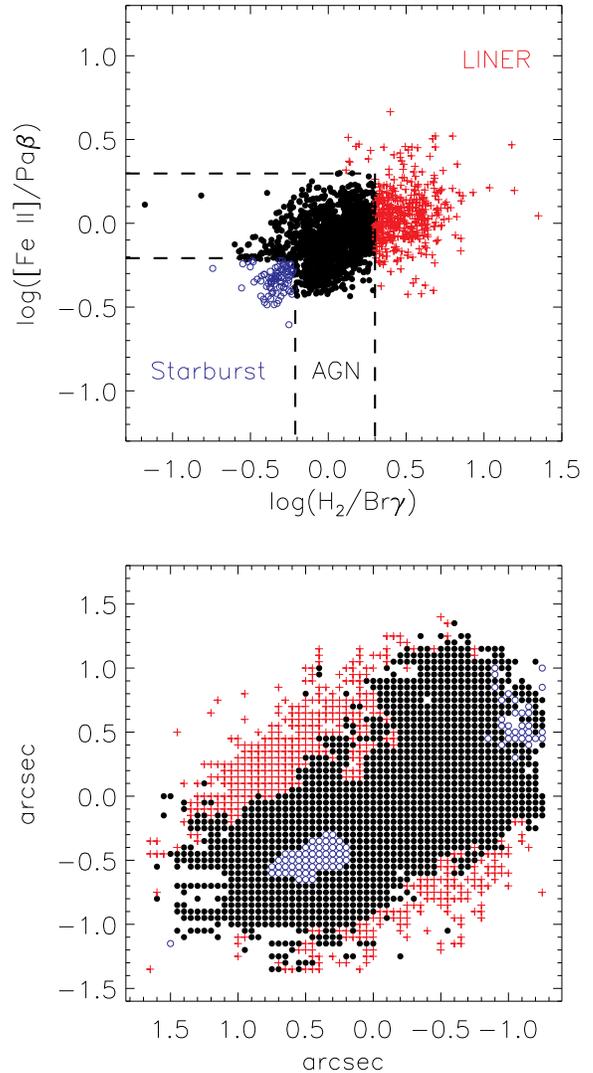}
\caption{Top:\feii$\lambda1.25\mu$m/\pb~$vs$~\h2$\lambda2.12\mu$m/\br~ line-ratio diagnostic diagram. The dashed lines delimit regions with ratios typical of 
Starbursts (blue-open circles), Seyferts (black-filled circles) and LINERs (red crosses). Bottom: Spatial position of each point from the diagnostic 
diagram.}
\label{diagn}
\end{figure}

In order to investigate the nature of the circum-nuclear line-emitting region we constructed the spatially resolved 
spectral-diagnostic diagram \feii$\lambda1.25\mu$m/\pb~$vs$~\h2$\lambda2.12\mu$m/\br, originally proposed for 
non-resolved spectra \citep{larkin98,rodriguez-ardila04,rodriguez-ardila05a}. Typical values for the nucleus of Seyfert 
galaxies are   0.6$\lesssim$\feii/\pb$\lesssim$2.0 and 0.6$\lesssim$\h2/\br$\lesssim$2.0 
\citep[e.g.][]{rodriguez-ardila05a}. The diagram is shown in the top panel of Fig.\,\ref{diagn}, in which black filled circles represent 
typical values for Seyfert galaxies, blue open circles for Starbursts and red crosses  for LINERs.
Most ratios present Seyfert-like, although some points are observed in the LINER and Starburst regions of the diagnostic diagram. 
\citet{stoklasova09} analyzed optical IFU data for Mrk\,1066 and presented several optical diagnostic diagrams. A comparison 
between the optical and the present near-IR diagram shows that they are consistent with each other.

The locations of the distinct line ratios are shown in the bottom panel of Fig.~\ref{diagn}. Seyfert-like line ratios are observed approximately along the 
radio jet and ionization cone \citep[PA$\approx135/315^\circ$,][]{bower95,nagar99} suggesting that the line emission in these regions is 
driven by the AGN. The Starburst ratios are  mainly observed in a region  at 0\farcs5 south-east of the nucleus 
at the  position of the secondary emission peak of the H recombination lines, as described in the Sec.\,\ref{disc_distr}. 
Approximately at this same position there is also a higher flux in the continuum (see Fig.\,\ref{cnt}).  
Fig.\,\ref{diagn}
 shows also some Starburst ratios at 1$^{\prime\prime}$ north-west in a region approximately coincident  with a 
small enhancement of the H emission (see Fig.\,\ref{oned}) and with a knot in the optical continuum image (see Fig.~\ref{cnt}).  We interpret the Starburst 
ratios as originating in star formation regions (SFRs) located at these positions. 
This conclusion is supported by the observations of knots in the optical continuum image and its absence in the near-IR continuum, as young stars 
contribute mostly to the optical emission, while in the near-IR the continuum is dominated by the emission from old stars. 
%We can estimate the age of these SFRs using the evolutionary model calculated by \citet{davies07} together with the observed 
%\br\ equilvalent width of $\approx30$\,\AA\ for both regions. The resulting age depends on the star formation regime and is
% in the range from 10\,Myr for an instantaneous burst to 200\,Myr for contiuous star formation. This range is in good agreement 
%with the ones observed for other Seyfert galaxies \citep{davies07}. 
At distances $\gtrsim$0\farcs5 perpendicular to the orientation of the ionization cone the \h2$\lambda2.12\mu$m/\br~ratio reaches the highest values, 
being in the region of LINERs in the diagnostic diagram. We interpret this low ionization region as being originated by a diffuse radiation field 
 escaping through the walls of the ionization cone, which have enough energy to excite the \h2.
 A similar scenario is observed for the well-studied Seyfert galaxy NGC\,4151 \citep{kraemer08} 
and attributed to the origin of the near-IR lines in regions away from its ionization cone \citep{storchi-bergmann09}.
%This result is consistent with that of \citet{stoklasova09}, who found that the  
%[O\,{\sc i}]$\,\lambda6300$/\ha~and \nii$\,\lambda6583$/\ha~ratios are also higher at these positions, whit values typical of LINERs. We interpret this result as 
%due to additional \h2~emission excited by a difuse X-ray radiation (see next section).   

\subsection{H$_2$ excitation} \label{disc_h2}

\begin{figure*}
\centering
\includegraphics[scale=0.95]{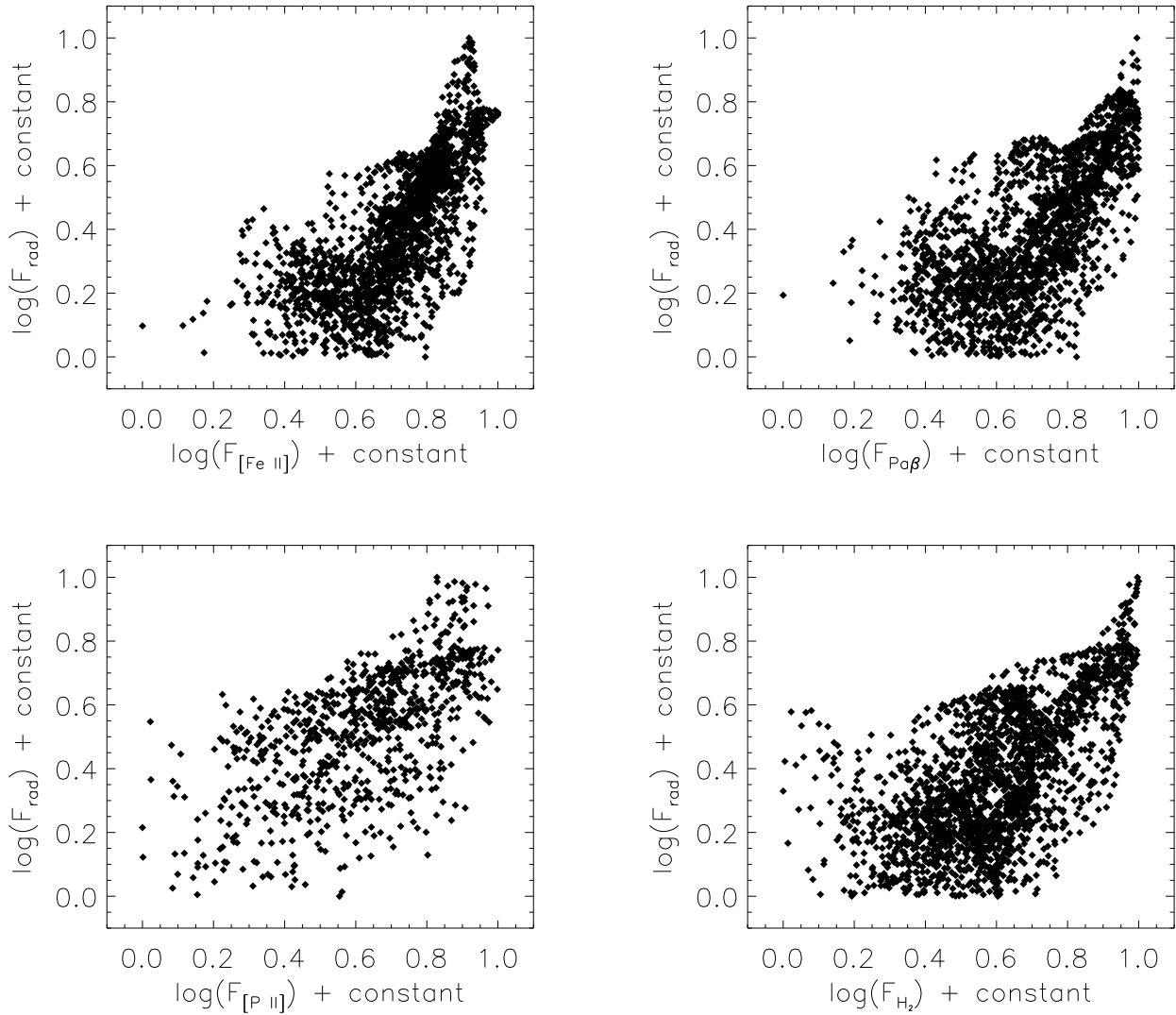}
\caption{Correlation between emission-line fluxes (top left: \feii$\,\lambda1.2570\,\mu$m, top right:  \pb, bottom
 left: \pii$\,\lambda1.1886\,\mu$m and bottom left: \h2$\,\lambda 2.1218\,\mu$m) and 3.6~cm radio emission.}
\label{correlation}
\end{figure*}

%The H$_2$ lines can be excited in two different ways: fluorescent excitation  \citep{black87} and collisional excitation produced by shocks \citep{hollenbach89}, X-ray heating \citep{maloney96} or  photodissociation by UV photons. The method commonly used to differentiate collisional excitation from fluorescence is based on the flux ratios of H$_2$ lines in the {\it K} band. Based on measurements of such ratios for a large sample of galaxies,  \citet{rodriguez-ardila05} concluded that  fluorescence is not important for AGNs supporting previous studies by \citet{veilleux97}, \citet{moorwood90} and \citet{fischer87}. 

The molecular hydrogen flux distribution and kinematics in the central regions of active galaxies have 
been the subject of several recent studies. Nevertheless, its excitation mechanisms is still in 
debate \citep[e.g.][]{reunanen02,rodriguez-ardila04,rodriguez-ardila05a,davies05,riffel06,riffel08,sanchez09, ramos-almeida09, 
hicks09,storchi-bergmann09}. 
The H$_2$ emission lines can be excited by two mechanisms: (i) fluorescent
excitation through absorption of soft-UV photons (912--1108 \AA) in
the Lyman and Werner bands -- present both in star-forming regions and surrounding AGNs
\citep{black87} and (ii) collisional excitation (usually referred to as thermal processes) due to heating of the gas 
by shocks, due to interaction of a radio jet with the interstellar medium \citep{hollenbach89} 
or by X-rays from the central AGN  \citep{maloney96}. 

We can use the \h2$\lambda$2.1218/\br\ emission-line ratio to investigate the origin of  the H$_2$ emission. 
 Typical values for Starburst galaxies, where the main heating agent is the UV radiation, are in the range 
\h2/\br$\lesssim$0.6 \citep{rodriguez-ardila04,rodriguez-ardila05a}, while for
Seyferts this ratio is larger because of additional thermal
excitation by shocks or X-rays from the AGN. As observed in Figs.~\ref{ratio} and \ref{diagn}, Mrk\,1066 presents two regions with
 \h2$\lambda$2.1218/\br\ $<$ 0.6. The first located at $\sim1^{\prime\prime}$\,north-west from the nucleus and the second 
at  $\approx$0\farcs5\,south-east. As discussed in Sec.\,\ref{disc_diagn}, we interpret this result as due to an enhancement of
 \br~due to ionization from young stars in SFRs at these locations. In regions away from these SFRs, \h2$\lambda$2.1218/\br\ is larger
 than 0.6 reaching $\sim2.6$  in regions more than 0\farcs7 from the nucleus in the direction perpendicular to the radio jet, suggesting additional H$_2$ emission
 due to thermal processes.

Another line ratio which can be used to study the \h2\ excitation is \h2$\lambda$2.2477/$\lambda$2.1218. 
For fluorescent excitation, typical values are $\sim0.55$, while for thermal processes this ratio is 
 $\sim0.1-0.2$ \citep[e.g.][]{mouri94,reunanen02,rodriguez-ardila04,storchi-bergmann09}.
In the case of active galaxies, the  fluorescent process seems not to be important \citep[e.g.][]{rodriguez-ardila04,rodriguez-ardila05a,riffel06}. 
 As observed in Tab.\,\ref{fluxes}, \h2$\lambda$2.2477/$\lambda$2.1218$\approx$0.1 in all positions, indicating  
that the  \h2 emission in most of the circum-nuclear region of Mrk\,1066 is also due to excitation by thermal processes.

\begin{figure}
\centering
\includegraphics[scale=0.45]{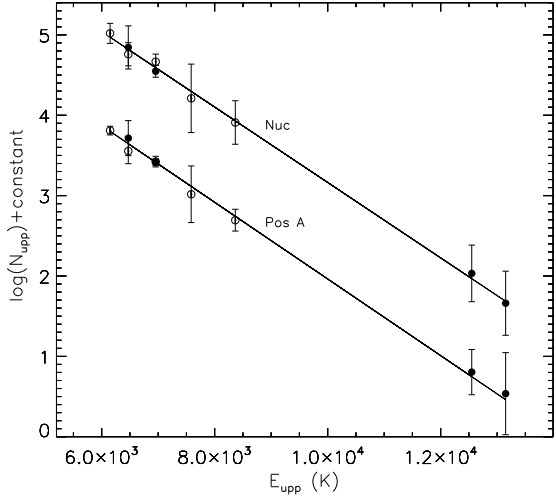}
\caption{Relation between $N_{\rm upp} = \frac{F_i \lambda_i}{A_i g_i}$ and $E_{\rm upp}=T_i$ for the H$_2$ 
emission lines for thermal excitation at the nuclear position and at 0\farcs5\,north-west 
from the nucleus. {\it Ortho} transitions are shown as filled circles and {\it para} transitions
 as  open circles.
From left to right the transitions shown are: 1--0\,Q(1), 1--0\,S(0), 1--0\,Q(2), 1--0\,S(1), 1--0\,Q(4),
 1--0\,Q(5), 2--1\,S(1) and 1--0\,S(2)
}
\label{h2_temp}
\end{figure}

We can also use the observed fluxes for all \h2\ emission lines in the K band to calculate the thermal excitation temperature.
 Following \citet{wilman05}, we have investigated the expression \citep[see also][]{storchi-bergmann09}:

\begin{equation}
 {\rm log}\left(\frac{F_i \lambda_i}{A_i g_i}\right)={\rm constant}-\frac{T_i}{T_{\rm exc}},
\end{equation}
where $F_i$ is the flux of the $i^{th}$ H$_2$ line, $\lambda_i$ is its wavelength, 
$A_i$ is the spontaneous emission coefficient, $g_i$ is the statistical                         
weight of the upper level of the transition, $T_i$ is the
energy of the level expressed as a temperature and $T_{\rm exc}$ is the excitation temperature.
 This relation is valid for thermal excitation, under the assumption of an
{\it ortho:para} abundance ratio of 3:1. In Fig.\,\ref{h2_temp} we present the observed values for 
$N_{\rm upp}=\frac{F_i \lambda_i}{A_i g_i}$ (plus an arbitrary constant) $vs$ $E_{\rm upp}={T_i}$ for the nuclear spectrum 
and for an extra-nuclear one at 0\farcs5~north-west from the nucleus (position A in Fig.\,\ref{espectro}). 
The resulting fit of the above relation is shown in Fig.\,\ref{h2_temp} as a continuum line and resulted in 
an excitation temperature of  $T_{\rm exc}=2131\pm40$\,K for the nucleus and 
$T_{\rm exc}=2097\pm40$\,K  for the position A. %The fact that
%all emission lines are on the relation confirms that the H$_2$ is
%in thermal equilibrium at $T_{\rm exc}$, and that the rotational and
%vibrational temperatures are the same, ruling out a significant contribution from 
%fluorescence.
 The fact that the {\it ortho} lines (1-0 S(1), 2-1 S(1), 1-0 Q(1), etc) 
give the same result as the {\it para} lines (1-0 S(0), Q(2), Q(4), etc) confirm that the H$_2$ is
in thermal equilibrium at $T_{\rm exc}$ and that the {\it ortho} to {\it para} ratio is 3 as assumed. This is also what is
expected for thermal equilibrium.

 The heating of the \h2\  can be provided  
by nuclear X-rays and/or by shocks due to the interaction of the radio jet with the interstellar 
medium. Nevertheless distinguishing between both mechanisms is usually a hard task. 
Correlations between the radio or X-ray emission with the H$_2$ emission could be useful to 
investigate the importance of these mechanisms. \citet{quillen99}, using HST, 
imaged a sample of 10 Seyfert galaxies in H$_2$ and looked for correlations with radio 6~cm and
hard X-ray flux. They found no correlation with X-rays  and  a weak
correlation with radio 6-cm, suggesting that no single mechanism is
likely to be responsible for the \h2\ excitation in Seyfert galaxies.
 
In Fig.\,\ref{correlation} we present plots for the  \feii$\,\lambda1.2570\mu$m (top left panel), 
\pii$\,\lambda1.1886\mu$m (bottom left), \pb~(top right) and H$_2\,\lambda2.1218\mu$m (bottom right) emission-line fluxes
 {\it versus} the radio flux at 3.6\,cm for each spatial pixel. As observed in this figure the  flux distributions correlate 
very well with the radio emission. We used the routine {\sc r\_correlate}  in {\sc idl}\footnote{http://ittvis.com/idl} programing language to obtain the 
Sepearman correlation coefficient ($R$). The best correlation is found for \feii~with $R=0.77$, followed by \pb~with 
$R=0.70$, \pii~with $R=0.64$ and H$_2$ with $R=0.62$. These correlation can be understood in two ways: (i) the interaction 
between the radio jet and the ISM produces the heating necessary to collisionally excitation the emitting gas and/or (ii) 
the radio jet only compresses the gas enhancing its emission, excited by other mechanism. The fact that 
 \pb~emission shows a better correlation with the radio  than the H$_2$ emission suggests that the radio jet only compresses
 the gas, since the H recombination lines are not sensible to collisional excitation. Nevertheless, 
shocks due to the radio jet cannot be completely discarded since the ionized and molecular hydrogen emission could arise from 
distinct locations along the line-of-sight.

Following \citet{rodriguez-ardila04} and \citet{riffel08} we evaluate the emergent flux for the \h2$\,\lambda2.1218\,\mu$m 
of a gas cloud illuminated by a source of hard X-rays using the model  of \citet{maloney96}. 
\citet{shu07} obtained a 2--10\,keV X-ray flux of  $F_X\sim 2.3\times10^{-13}\,{\rm erg s^{-1} cm^{-2}}$ and a attenuating column density
 of $N_{H} > 10^{24}\,{\rm cm^{-2}}$ for Mrk\,1066 using Chandra observations. The $N_{H}$ value is measured along the line-of-sight and according to 
the unified model $N_{H}$ may be smaller in other directions, since Mrk\,1066 has a Seyfert 2 nucleus and thus 
the dusty torus is seen edge-on, enhancing the column density only along the line-of-sight \citep{antonucci93}. The column density in the \h2\ emitting 
region can be estimated using  $N_H\approx5.2\times10^{21}\,E(B-V)$ [cm$^{-2}$] \citep{shull85}. Using the $E(B-V)$ values from Fig.~\ref{ratio} we obtain  $N_H\approx 10^{21}$~ cm$^{-2}$, which is in good agreement with the value obtained by  \citet{guainazzi05} -- $N_{H} = 1.21\times10^{21}\,{\rm cm^{-2}}$ -- 
 using XMM-Newton X-ray observations. The lower value of $N_H$ obtained with the XMM-Newton relative to the one with  Chandra data may be due to the fact that 
former has a larger aperture than latter, diluting the effect of the nuclear dust.  Using $N_H\approx10^{21}$~ cm$^{-2}$ as the appropriate value for the 
emitting gas, 
the emergent fluxes calculated from the X-ray excitation model can account for the \h2~emission in most locations of the observed field of Mrk\,1066.
% Nevertheless, this result should be taken with caution due to the high discrepancies between the measured hard X-ray fluxes and column densities.

%In order to test if X-ray emission can account for the observed H$_2$ emission in Mrk\,1066 we calculate the emergent h$_2$ flux of a gas cloud
% illuminated by a source of hard X-rays with intrinsic luminosity $L_X$ from the model of X-ray excitation of \citet{maloney96}. 
% Following \citet{rodriguez-ardila04} and \citet{riffel08}, the effective ionization parameter $\xi_{\rm eff}$ can be obtained by
%\begin{equation}
%\xi_{\rm eff}\approx{\rm 100}\frac{L_X}{n_5 d^2 N_{22}^{0.9}},
%\end{equation}
%where $d$ is the distance from the X-ray source in parsecs, $n_5$[10$^{-5}$cm$^{-3}$]
% is the total hydrogen gas density, and $N_{22}$[10$^{22}$cm$^2$]  is the attenuating column density between the AGN and the 
%emitting cloud. Using  the 2--10~KeV luminosity of $L_X=4\pi d^2\,F_X=?????$,
% where $F_X\sim 3.6\times10^{-13}\,{\rm erg s^{-1} cm^{-2}}S$ is the hard X-ray flux,
% and column density of $N_{22} = 1.21\times10^{21}\,{\rm cm^{-3}}$ from \citep{guainazzi05} we obtained 
%the emergent fluxes for distances of 50, 100, 150, 200, 250 and 300~pc from the nucleus. These fluxes are 
%shown in Table\,\ref{Xflux} for an aperture of 0\farcs25$\times$0\farcs25. From this table we obtain that the 
%X-ray excitation can account for at least {\bf XXXXXXXXXXXXXXXXXX\% }. 

The line H$_2$ 2--1\,S(3) ($\lambda 2.0649\mu$m) can also be used to investigate the H$_2$ excitation. In the case of X-ray irradiated gas this 
line is expected to be absent \citep{davies05}. Our spectral range does not include this line, but \citet{rogerio06} 
obtained a near-IR nuclear spectrum for an aperture of 1\farcs6$\times$0\farcs6 in which the H$_2$ 2--1\,S(3) emission line is not 
detected, supporting the above conclusion that X-ray heating contribute to the observed H$_2$ emission.

From the discussion above, we conclude that heating by  X-rays from the central AGN may be the dominant excitation mechanism of the \h2. This 
conclusion is in good agreement with those obtained for other active galaxies \citep[e.g.][]{storchi-bergmann99,riffel08,storchi-bergmann09}.
 Nevertheless, as discussed above, some contribution of the radio jet to the \h2~excitation cannot be discarded. In fact, the interaction
 of the radio jet with the circum-nuclear gas seems to be important in the excitation of the \h2~in some galaxies, such as ESO\,428-G14 
\citep{riffel06}.

%\begin{table*}
%\caption{Comparasion of the observed H$_2\,\lambda2.1218\,\mu$m fluxes and calculated using models of \citet{maloney96} for an aperture of 0\farcs25$\times$0\farcs25 for hydrogen densities $n=10^5$\,cm$^{-3}$ and $n=10^3$\,cm$^{-3}$ .}
%\centering
%\begin{tabular}{c c c c c c }
%\hline
%  & \multicolumn{3}{c}{Observed} & \multicolumn{2}{c}{Predicted}     \\
%      & PA=315$\degr$         & PA=135$\degr$          &  PA=45$\degr$          &  \multicolumn{2}{c}{$\rm n=10^5\,cm^{-3}$}  \\
%\hline
%$d$   & \small{log($F_{H_2}$)}& \small{log($F_{H_2}$)} & \small{log($F_{H_2}$)} & log($\xi_{\rm eff})$ & log($F_{H_2}$)\\
%50    &  -15.0                &   -15.0 	       &    -15.1               & -3.6                 & -12.4          \\
%100   &  -15.0 		      &   -15.2		       &    -15.3               & -4.2                 & -12.4          \\
%150   &  -15.0                &   -15.3 	       &    -15.6               & -4.5                 & -12.5           \\
%200   &  -15.2                &   -15.6		       &    -15.7               & -4.8                 & -12.5           \\
%250   &  -15.7                &   -15.7		       &    -15.8               & -5.0                 & -12.5           \\
%300   &  -16.0                &   -15.8		       &    -16.0               & -3.9                 & -16.7           \\
%\hline
%\end{tabular}
%\label{Xflux}
%\end{table*}

\subsection{[Fe\,{\sc ii}] excitation}

The \feii$\,\lambda1.2570\mu$m/\pb~line ratio is controlled by the ratio 
between the volumes of partially to fully ionized regions, as the \feii~emission is excited in partially
 ionized gas regions. In AGNs, such regions can be created by X-ray \citep[e.g.][]{simpson96} and/or shock 
\citep[e.g.][]{forbes93} heating of the gas.  

The \feii$\,\lambda1.2570\mu$m/\pb~line ratio can be used to investigate the excitation mechanism 
of the \feii~emission. For Starburst galaxies,  \feii/\pb$\lesssim0.6$ and 
for supernovae for which  shocks are the main excitation mechanism, 
this ratio has values larger than 2 \citep{rodriguez-ardila04,rodriguez-ardila05a}. Thus this ratio 
can be used to measure the relative contribution of photoionization and shocks, with values 
close to 0.6 meaning that photoionization dominates, while values close to 2 that
 shock excitation is dominant. As observed in Figs.\,\ref{ratio} and \ref{diagn}, for Mrk\,1066 
typical values for this ratio are 0.6--1.5, although in the regions co-spatial with the SFRs 
\feii/\pb$\lesssim0.6$ due to an enhancement of the \pb~emission. The highest values are 
observed in locations  close to the edge of the radio structure suggesting that the excitation due to shocks 
by the radio jet is important in these regions. The stronger correlation between the \feii~and the radio emission than those 
observed for the other emission lines (see Fig.\,\ref{correlation}) support a higher contribution of the radio jet for the 
excitation of  \feii~than for  \h2. Radio shocks have been appointed  as the main \feii~excitation mechanism
 for some Seyfert galaxies such as ESO\,428-G14 \citep{riffel06} and being not negligible in others, 
as in NGC\,4151 \citep{storchi-bergmann09}.

Another tracer of the origin of the \feii~emission is the 
line ratio [Fe\,{\sc ii}]$\lambda$1.2570$\,\mu$m/[P\,{\sc ii}]$\lambda$1.8861$\,\mu$m. 
These two lines have similar excitation  temperatures, and their parent ions
have similar ionization potentials and radiative recombination coefficients. 
As pointed out in \citet{oliva01} -- see also \citet{storchi-bergmann09} -- values 
larger than 2  indicate that shocks have passed
through the gas destroying the dust grains, releasing the
Fe and thus enhancing its observed abundance. For supernova remnants 
where shocks are the dominant excitation mechanism [Fe\,{\sc ii}]/[P\,{\sc ii}]
 is typically  higher than 20 \citep{oliva01}. As observed in Fig.\,\ref{ratio}, Mrk\,1066 presents \feii/\pii$\sim$3
 at most locations, but in some regions close to the borders of the radio structure \feii/\pii~reaches values up to 9.5, 
indicating that shocks due to the radio jet are important in these locations in agreement with 
the highest values obtained for the \feii/\pb\ at the same locations. %Nevertheless, the 
%\feii/\pii~line ratio is never as high as $\sim20$, which indicates that the main excitation mechanism is the X-ray heating.  

From the discussion above, we  conclude that X-ray heating may be the dominant excitation mechanism of the \feii\ in Mrk\,1066, exceptions are the regions 0\farcs85 
north-west of the nucleus where there is a radio hot spot and in the borders of the radio structure, where shocks seem to play a more 
important role. In any case, the contribution of shocks due to the radio jet seems to be more important for the \feii~emission than for the \h2~emission, as 
indicated by the better correlation observed for with the radio emission and the \feii~than the one observed for the \h2. 
This result is in good agreement with the ones obtained for other galaxies \citep[e.g.][]{rodriguez-ardila04,
rodriguez-ardila05a,riffel06,storchi-bergmann99,storchi-bergmann09}.

%The relative importance of X-rays in the \feii excitation can be also investigated using the excitation models 
%of \citet{maloney96}, as done in the last section for the \h2 emission. In the Table\,\ref{Xflux} we present the observed 
%and predicted fluxes for the \feii$\,\lambda1.2570\,\mu$m. As can be observed in this table, 
%X-rays can account for $\sim${\bf XXXXXXXXX\%} of the \feii observed emission. Nevertheless, as discussed for the \h2, the X-ray flux 
%may be higher in regions perpendicular to the dusty torus, since it may absorbs part of the X-ray emission in the line of sight.

\subsection{Mass of ionized and molecular gas}

Following \citet{riffel08} and \citet{storchi-bergmann09} we  estimate the mass of ionized gas in the inner 
700\,$\times$\,700\,pc$^2$ by 
\begin{equation}
 M_{HII}\approx3\times10^{17}\left(\frac{F_{\rm Br\gamma}}{\rm erg\,s^{-1}cm^{-2}}\right)\left(\frac{d}{\rm Mpc}\right)^2 ~~~~~~[{\rm M_\odot}],
\end{equation}
where  $F_{\rm Br\gamma}$ is the integrated flux for the \br~emission line and $d$ is the distance to Mrk\,1066. 
We have assumed an electron temperature $T=10^4$\,K and electron density $N_e=100\,{\rm cm^{-3}}$. 

 The mass of molecular gas can 
be obtained as
\begin{equation}
 M_{H_2}\approx5.0776\times10^{13}\left(\frac{F_{H_{2}\lambda2.1218}}{\rm erg\,s^{-1}\,cm^{-2}}\right)\left(\frac{d}{\rm Mpc}\right)^2~~~~~~[{\rm M_\odot}],
\label{mh2}
\end{equation}
where  $F_{H_{2}\lambda2.1218}$ is the integrated flux for the \h2$\,\lambda2.1218\,\mu$m emission line
and we have used the vibrational temperature T=2000\,K we have obtained in Sec.~\ref{disc_h2}.

Integrating over the whole IFU field we obtain $F_{\rm Br\gamma}\approx2.5\times10^{-14}\,{\rm erg\,s^{-1}\,cm^{-2}}$ and
 $F_{H_{2}\lambda2.1218}\approx2.8\times10^{-14}\,{\rm erg\,s^{-1}\,cm^{-2}}$,  resulting in   $M_{HII}\approx1.7\times10^7\,{\rm M_\odot}$
  and $M_{H_2}\approx3.3\times10^3\,{\rm M_\odot}$.  The mass of molecular gas is 10$^4$ times smaller than of ionized gas but, 
as discussed in \citet{storchi-bergmann09}, this 
H$_2$ mass represents only that of hot gas emitting in the near-IR, heated by X-rays and maybe also by shocks from the central 
AGN.  The total mass in molecular gas (including the cold gas which does not emit in the near-IR) is usually 10$^{5}-$10$^{7}$ times that in hot \h2\ 
\citep{dale05} suggesting that the total molecular gas mass is at least 10$^{8}\,{\rm M_\odot}$.

\section{Conclusions}

We used integral field J- and K$_{\rm l}$-Band spectroscopy of the inner $700\times700$\,pc$^2$ of the Seyfert galaxy Mrk\,1066, obtained with the
Gemini NIFS at spatial resolution $\sim$35\,pc and spectral resolution $\sim$3.3\,$\AA$, to map the near-IR continuum and emission-line flux
distributions, as well as the properties of the nuclear spectrum. Our main conclusions are:

\begin{itemize}
\item The nucleus contains an unresolved infrared source whose continuum is well reproduced by the emission of a dust structure with temperature 
$\sim$830\,K and mass $\sim1.4\times10^{-2}\,{\rm M_\odot}$ which we identify as the dusty torus of the Unified Model;

\item Emission-line fluxes (except for the coronal lines) are elongated in PA$=135/315^\circ$ in agreement with
previous optical \oiii\ imaging and also following the radio jet extending beyond the IFU field. Except for the H lines, the emission
is stronger to the north-west than to the south-EAST in association with the radio hot spot implying that at
least part of the emitting gas is co-spatial with the radio outflow. The H emission is stronger to the south-east, where we find a large
region of star-formation.

\item There is a strong correlation between the radio emission and the highest emission-line fluxes suggesting that the radio jet plays
an important role for these intensity levels. At lower line flux values this correlation  disappears  indicating a
contribution from gas in the galactic plane.

\item The coronal line emission in [Ca\,{\sc viii}]$\,\lambda2.3220\,\mu$m and [S\,{\sc ix}]$\,\lambda1.2524\,\mu$m
are unresolved by our observations indicating that they
originate within a radius of 18\,pc from the nucleus.

\item The \feii/\pb~$vs$~\h2/\br\ line-ratio diagnostic diagram presents typical
values for Seyfert nuclei along the ionization cone indicating that the line emission is powered by the AGN in these regions. Some radiation
escapes through the walls of the ionization cone and powers a low ionization emission-line region in locations perpendicular to the
cone and radio jet.  At the positions  0\farcs5 south-east  and at 1$^{\prime\prime}$ north-west of the nucleus
Starburst-like values are observed due to additional emission of the H lines from star forming regions.

\item The reddening map obtained via the \pb/\br~line ratio present a S-shaped structure with high values oriented along the
PA$\approx$135/315$^\circ$, reaching up to $E(B-V)=1.7$. Lower values are observed in regions away from this structure with
typical values of  $E(B-V)=0.5$. The reddening in the continuum is smaller with typical values of $E(B-V)=0.9$ along the
S-shaped structure. 

\item  The \h2\ gas has an excitation temperature $T_{\rm exc}\approx2100$\,K and its emission is mainly due to X-ray excitation from
the central AGN, which is also the main excitation mechanism of the \feii\ emission. Emission due to shocks produced
by the radio jet may contribute a small fraction to the emission of both lines, with a larger contribution for the \feii\ line.
 
\item The mass of ionized gas in the inner $700\times700$\,pc$^2$ of Mrk\,1066 is $M_{HII}\approx1.7\times10^7\,{\rm M_\odot}$,
while that of hot molecular
gas is $M_{H_2}\approx3.3\times10^3\,{\rm M_\odot}$.

\end{itemize}

%\begin{table}
%\caption{Kinematic PA for differente emission lines}
%\centering
%\begin{tabular}{c c c c c}
%H$_2$ & Pa$\beta$ & [Fe\,{\sc ii}] &  [P\,{\sc ii}] & Stars \\
%\hline 
%129$\degr$ & 115$\degr$     & 119.5$\degr$   & 125.5$\degr$          & 121$\degr$ \\
%\hline
%\end{tabular}
%\label{fluxes}
%\end{table}

\section*{Acknowledgments}
Based on observations obtained at the Gemini Observatory, 
which is operated by the Association of Universities for Research in Astronomy, Inc., under a cooperative agreement with the 
NSF on behalf of the Gemini partnership: the National Science Foundation (United States), the Science and Technology 
Facilities Council (United Kingdom), the National Research Council (Canada), CONICYT (Chile), the Australian Research 
Council (Australia), Minist\'erio da Ci\^encia e Tecnologia (Brazil) and south-eastCYT (Argentina).  
This research has made use of the NASA/IPAC Extragalactic Database (NED) which is operated by the Jet
 Propulsion Laboratory, California Institute of  Technology, under contract with the National Aeronautics and Space Administration.
This work has been partially supported by the Brazilian institution CNPq.

\label{lastpage}


\begin{thebibliography}{99}

\bibitem[\protect\citeauthoryear{Antonucci}{1993}]{antonucci93} Antonucci, R. 1993, ARA\&A, 31, 473.

\bibitem[\protect\citeauthoryear{Barbosa et al.}{2009}]{barbosa09} Barbosa, F. K. B., Storchi-Bergmann, T., Cid Fernandes, R., Winge, C., Schmitt, H., 2009, MNRAS, 396, 2.

\bibitem[\protect\citeauthoryear{Barvainis}{1987}]{barvainis87} Barvainis, R. 1987, ApJ, 320, 537

\bibitem[\protect\citeauthoryear{Black \& van Dishoeck}{1987}]{black87} Black, J. H., \& van Dishoeck, E. F.  1987, ApJ, 322, 412.

%\bibitem[\protect\citeauthoryear{Boone et al.}{2007}]{boone07} Boone, F., Baker, A. J., Schinnerer, E., Combes, F., Garc\'\i a-Burillo, S.,
% Neri, R., Hunt, L. K., L\'eon, S., Krips, M., Tacconi, L. J., \& Eckart, A., 2007, A\&A, 471, 113.

\bibitem[\protect\citeauthoryear{Bower et al.}{1995}]{bower95} Bower, G., Wilson, A., Morse, J. A., Gelderman, R., Whitle, M., \& Mulchaey, J., 1995, ApJ, 454, 106.

\bibitem[\protect\citeauthoryear{Cardelli, Clayton \& Mathis}{1989}]{cardelli89} Cardelli,J. A., Clayton, G. C. \& Mathis, J. S., 1989, ApJ, 345,245.

\bibitem[\protect\citeauthoryear{Dale et al.}{2005}]{dale05} Dale, D. A., Sheth, K., Helou, G., Regan, M. W., \& H\"uttemeister, S., 2005, ApJ, 129, 2197.

\bibitem[\protect\citeauthoryear{Davies et al.}{2005}]{davies05} Davies, R. I., I., Sternberg, A., Lehnert, M. D., \& 
 Tacconi-Garman, L. E., 2005, ApJ, 633, 105. 

\bibitem[\protect\citeauthoryear{Davies et al.}{2007}]{davies07} Davies, R. I., S\'anchez, F. M., Genzel, R., Tacconi, L. J., Hicks, E. K. S.,
 Friedrich, S., \& Sternberg, A. ApJ, 671, 1388.

\bibitem[\protect\citeauthoryear{Elitzur}{2008}]{elitzur08} Elitzur, M. 2008, NewAR, 52, 274

\bibitem[\protect\citeauthoryear{Forbes \& Ward}{1993}]{forbes93} Forbes, D. A. \& Ward, M. J.  1993, ApJ, 416, 150. 


%\bibitem[\protect\citeauthoryear{Garc\'\i a-Burillo et al.}{2005}]{garcia-burillo05} Garcia-Burillo, S., Combes, F., Schinnerer, E., 
%Boone, F., \& Hunt, L. K., 2005, A\&A, 441, 1011.


\bibitem[\protect\citeauthoryear{Granato \& Danese}{1994}]{granato94} Granato, G. L., \& Danese, L. 1994, MNRAS, 268, 235

\bibitem[\protect\citeauthoryear{Guainazzi, Matt \& Perola}{2005}]{guainazzi05} Guainazzi, M., Matt, G., \& Perola, G. C., 2005, A\&A, 444, 119.  

\bibitem[\protect\citeauthoryear{Hollenbach \& McKee}{1989}]{hollenbach89} Hollenbach, D., \& McKee, C. F., 1989, ApJ, 342, 306.

\bibitem[\protect\citeauthoryear{Hicks et al.}{2009}]{hicks09} Hicks, E. K. S., Davies, R. I., Malkan, M. A.,
 Genzel, R., Tacconi, L. J.; S\'anchez, F. M., Sternberg, A., 2009, ApJ, 696, 448. 

\bibitem[\protect\citeauthoryear{Kraemer, Schmitt \& Crenshaw}{2008}]{kraemer08}  Kraemer S. B., Schmitt H. R., \& Crenshaw D. M., 2008, ApJ, 679, 1128.

\bibitem[\protect\citeauthoryear{Kishimoto et al.}{2007}]{kishimoto07} Kishimoto, M., H\"onig, S. F., Beckert, T., \& Weiglt, G., 2007, 
A\&A, 476, 713.

\bibitem[\protect\citeauthoryear{Knop et al.}{2001}]{knop01} Knop, R. A., Armus, L., Matthews, K., Murphy, T. W., \& Soifer, B. T., 2001, ApJ, 122, 764.


%\bibitem[\protect\citeauthoryear{Krips et al.}{2007}]{krips07} Krips, M., Neri, R., Garc\'\i a-Burillo, S., Combes, F., Schinnerer, E., Baker, A. J., Eckart, A., 
%Boone, F., Hunt, L., Leon, S., \& Tacconi, L. J., 2007, A\&A, 468, 63.

\bibitem[\protect\citeauthoryear{Larkin et al.}{1998}]{larkin98} Larkin, J. E., Armus, L., Knop, R. A., Soifer, B. T., \& Matthews, K., 1998, ApJS, 114, 59.

\bibitem[\protect\citeauthoryear{Lasker et al.}{1996}]{lasker96} Lasker, B. M., Doggett, J., McLean, B., Sturch, C., Djorgovski, S., de Carvalho, R. R., Reid, I. N., 1996, 
Astronomical Data Analysis Software and Systems V, A.S.P. Conference Series, George H. Jacoby and Jeannette Barnes, eds., 101, 88. 

	
\bibitem[\protect\citeauthoryear{Malkan, Gorjian \& Tam}{1998}]{malkan98} Malkan, M. A., Gorjian, V. \& Tam, R., 1998, ApJS, 117,25.

\bibitem[\protect\citeauthoryear{Maloney, Hollenbach \& Tielens}{1996}]{maloney96} Maloney, P. R.,  Hollenbach, D. J.,  \& Tielens, A. G. G. M., 1996, ApJ, 466, 561.

\bibitem[\protect\citeauthoryear{McGregor et al.}{2003}]{mcgregor03} McGregor, P. J. et al., 2003, Proceedings of the SPIE, 4841, 1581.

%\bibitem[\protect\citeauthoryear{Mouri et al.}{1990}]{mouri90} Mouri, H., Nishida, M., Taniguchi, Y., \& Kawara, K. 1990, ApJ, 360, 55.

%\bibitem[\protect\citeauthoryear{Mouri, Kawara \& Taniguchi}{1993}]{mouri93} Mouri, H., Kawara, K., \& Taniguchi, Y.  1993, ApJ, 406, 52.

\bibitem[\protect\citeauthoryear{Mouri}{1994}]{mouri94} Mouri, H. 1994, ApJ, 427, 777.

\bibitem[\protect\citeauthoryear{Mundell et al.}{2009}]{mundell09} Mundell, C.~G.,
Ferruit, P., Nagar, N., \& Wilson, A.~S.\ 2009, ApJ, 703, 802


\bibitem[\protect\citeauthoryear{Nagar et al.}{1999}]{nagar99}  Nagar, N. M., Wilson, A. S., Mulchaey, J. S. \& Gallimore, J. F., 1999, ApJS, 120, 209.

\bibitem[\protect\citeauthoryear{Osterbrock \& Ferland}{2006}]{osterbrock06} Osterbrock, D. E. \& Ferland, G. J., 2006, 
Astrophysics of Gaseous Nebulae and Active Galactic Nuclei, Second Edition, University Science Books, Mill Valley, California.


\bibitem[\protect\citeauthoryear{Oliva et al.}{2001}]{oliva01} Oliva, E. et al. 2001, A\&A, 369, L5.

\bibitem[\protect\citeauthoryear{Quillen et al.}{1999}]{quillen99} Quillen, A. C., Alonso-Herrero, A., Rieke, M. J., Rieke, G. H., Ruiz, M., \& Kulkarni, V. 1999, ApJ, 527, 696.

\bibitem[\protect\citeauthoryear{Ramos Almeida,   P\'erez Garc\'\i a \& Acosta-Pulido}{2009}]{ramos-almeida09} Ramos Almeida, C., P\'erez Garc\'\i a, A. M.,
 \& Acosta-Pulido, J. A., 2009, ApJ, 694, 1379.

\bibitem[\protect\citeauthoryear{Reunanen, Kotilainen \& Prieto}{2002}]{reunanen02} Reunanen, J., Kotilainen, J. K., \& Prieto, M. A., 2002, MNRAS, 331, 154. 

\bibitem[\protect\citeauthoryear{Rieke \& Lebofsky}{1981}]{rieke81} Rieke, G. H., \& Lebofsky, M. J., 1981, ApJ, 250, 87.

\bibitem[\protect\citeauthoryear{Riffel et al.}{2006}]{riffel06} Riffel, Rogemar A., Sorchi-Bergmann, T., Winge, C., Barbosa, F. K. B., 2006, MNRAS, 373, 2.

\bibitem[\protect\citeauthoryear{Riffel, Rodr\'\i guez-Ardila \& Pastoriza}{2006}]{rogerio06} Riffel, R., Rodr\'\i guez-Ardila, A., Pastoriza, M. G., 2006, A\&A, 457, 61.

\bibitem[\protect\citeauthoryear{Riffel et al.}{2008}]{riffel08} Riffel, Rogemar A., Storchi-Bergmann, T., Winge, C., McGregor, P. J., Beck, T., Schmitt, H. 2008, MNRAS, 385, 1129.

\bibitem[\protect\citeauthoryear{Riffel et al.}{2009a}]{riffel09a} Riffel, Rogemar A., Storchi-Bergmann, T., Dors, O. L., Winge, C., 2009, MNRAS, 393, 783.

\bibitem[\protect\citeauthoryear{Riffel et al.}{2009b}]{riffel09b} Riffel, Rogemar A., Storchi-Bergmann, T., McGregor, P. J., 2009, ApJ, 698, 1767.

\bibitem[\protect\citeauthoryear{Riffel et al.}{2009c}]{rogerio09} Riffel, R., Pastoriza, M. G., Rodr\'\i guez-Atdila, A., Bonatto, C. 
2009, accepted by MNRAS, arXiv:0907.4144


\bibitem[\protect\citeauthoryear{Rodr\'\i guez-Ardila et al.}{2004}]{rodriguez-ardila04} Rodr\'\i guez-Ardila, A.,  Pastoriza, M. G., Viegas, S., Sigut, T. A. A., \& Pradhan, A. K., 2004,  A\&A, 425, 457.

\bibitem[\protect\citeauthoryear{Rodr\'\i guez-Ardila, Riffel \& Pastoriza}{2005a}]{rodriguez-ardila05a} Rodr\'\i guez-Ardila, A., Riffel, R., \& Pastoriza, M. G. 2005,  MNRAS, 364, 1041.

\bibitem[\protect\citeauthoryear{Rodr\'\i guez-Ardila, Contini \& Viegas}{2005b}]{rodriguez-ardila05b} Rodr\'\i guez-Ardila, A.,  Contini, M., Viegas, S. 2005b  MNRAS, 357, 220

\bibitem[\protect\citeauthoryear{Rodr\'\i guez-Ardila \& Mazzalay}{2006}]{rodriguez-ardila06} Rodr\'\i guez-Ardila, A., \& Mazzalay, X. 2006,  MNRAS, 367, L57

\bibitem[\protect\citeauthoryear{S\'anchez et al.}{2009}]{sanchez09} S\'anchez, F. M., Davies, R. I., Genzel, R., Tacconi, L. J.,
 Eisenhauer, F., Hicks, E. K. S., Friedrich, S., \& Sternberg, A., 2009, ApJ, 691, 749.


\bibitem[\protect\citeauthoryear{Simpson et al.}{1996}]{simpson96} Simpson, C., Forbes, D. A., Baker, A. C., \& Ward, M. J. 1996, MNRAS, 283, 777.

\bibitem[\protect\citeauthoryear{Shu et al.}{2007}]{shu07} Shu, X. W., Wang, J. X., Jiang, P., Fan, L. L., Wang, T. G., 2007, ApJ, 657, 167.

\bibitem[\protect\citeauthoryear{Shull \& van Steenberg}{1985}]{shull85}  Shull, J. A. \& van Steenberg, M. E., 1985, ApJ, 294, 599.

\bibitem[\protect\citeauthoryear{Stoklasov\'a et al}{2009}]{stoklasova09} Stoklasov\'a, I., Ferruit, P., Emssellem, E., Jungwiert, B., P\'econtal, E., \&
 S\'anchez, S. F., 2009, A\&A, 500, 1287.

\bibitem[\protect\citeauthoryear{Storchi-Bergmann et al.}{1999}]{storchi-bergmann99} Storchi-Bergmann, T., Winge, C., Ward, M. J., Wilson, A. S., 1999, MNRAS, 304, 35.

\bibitem[\protect\citeauthoryear{Storchi-Bergmann et al.}{2009}]{storchi-bergmann09} Storchi-Bergmann, T., McGregor, P. Riffel,Rogemar A., Sim\~oes Lopes, R., 
Beck, T., Dopita, M., 2009, MNRAS, 394, 1148.

\bibitem[\protect\citeauthoryear{Ulvestad \& Wilson}{1989}]{ulvestad89} Ulvestad, J. S., \& Wilson, A. S., 1989, ApJ, 343, 659.

%\bibitem[\protect\citeauthoryear{van der Marel\& Franx}{1993}]{vandermarel93} van der Marel, R.P. \& Franx, M. 1993, ApJ, 407, 525 

\bibitem[\protect\citeauthoryear{Wilman, Edge \& Juhnstone}{2005}]{wilman05} Wilman, R. J., Edge, A. C., \& Juhnstone, R. M., 2005, MNRAS, 359, 755.


\end{thebibliography}
\end{document}